\documentclass[final,3p,times,twocolumn]{elsarticle}

\usepackage[T1]{fontenc}

\makeatletter
\def\ps@pprintTitle{%
 \let\@oddhead\@empty
 \let\@evenhead\@empty
 \let\@oddfoot\@empty
 \let\@evenfoot\@empty}
\makeatother

\usepackage{amsmath}
\usepackage{amssymb}
\usepackage{mathtools}

\usepackage{booktabs}
\usepackage{xcolor}
\usepackage{hyperref}
\usepackage{url}
\usepackage{xspace}

\usepackage{placeins} 

\usepackage{balance} 

\usepackage[capitalise,noabbrev]{cleveref}

\usepackage{adjustbox}

\usepackage{algorithm}
\usepackage[noEnd=true]{algpseudocodex}

\usepackage{siunitx}
\usepackage{physics}
\AtBeginDocument{\RenewCommandCopy\qty\SI}
\DeclareSIUnit\points{pts}
\DeclareSIUnit[]{\SIvoid}{{}}

\usepackage{todonotes}
\usepackage{ulem}

\DeclarePairedDelimiter\floor{\lfloor}{\rfloor}

\newcommand{\CC}{C\nolinebreak\hspace{-.05em}\raisebox{.4ex}{\tiny\bf +}\nolinebreak\hspace{-.10em}\raisebox{.4ex}{\tiny\bf +}}
\newcommand{\Cpp}{\CC}

\newcommand{\set}[1]{\left\{#1\right\}}
\newcommand{\tuple}[1]{\left(#1\right)}
\newcommand{\bigO}[1]{\mathcal{O}\left(#1\right)}

\newcommand{\dalesA}{\textit{5080\_54400}\xspace}
\newcommand{\dalesB}{\textit{5110\_54320}\xspace}
\newcommand{\dalesC}{\textit{5110\_54475}\xspace}
\newcommand{\dalesD}{\textit{5130\_54355}\xspace}

\newcommand{\sgRural}{\textit{sg28\_2}\xspace}
\newcommand{\sgUrban}{\textit{sg28\_4}\xspace}

\newcommand{\Hallway}{\textit{hallway6}\xspace}
\newcommand{\Speulderbos}{\textit{Speulderbos}\xspace}

\newcommand{\Paris}{\textit{Paris}\xspace}
\newcommand{\Lille}{\textit{Lille}\xspace}
\newcommand{\LilleTwo}{\textit{Lille 2}\xspace}

\usepackage{pgfplots}

\pgfplotsset{
  compat=1.18,
  log x ticks with fixed point/.style={
    xticklabel={
      \pgfkeys{/pgf/fpu=true}
      \pgfmathparse{exp(\tick)}%
      \pgfmathprintnumber[fixed relative, precision=3]{\pgfmathresult}
      \pgfkeys{/pgf/fpu=false}
    }
  },
  log y ticks with fixed point/.style={
    yticklabel={
      \pgfkeys{/pgf/fpu=true}
      \pgfmathparse{exp(\tick)}%
      \pgfmathprintnumber[fixed relative, precision=3]{\pgfmathresult}
      \pgfkeys{/pgf/fpu=false}
    }
  }
}

\usepackage{tikz}
\usetikzlibrary{arrows,patterns,positioning,shapes,fit,calc}

\usepackage[inline]{enumitem}
\newlist{compactenum}{enumerate}{1}
\setlist[compactenum,1]{nosep,label=\arabic*.}
\newlist{compactitem}{itemize}{1}
\setlist[compactitem,1]{nosep,label=\textbullet}
\newlist{inlineenum}{enumerate*}{1}
\setlist[inlineenum,1]{label=(\arabic*)}

\newcommand{\KNN}{\mbox{$k$-NN}\xspace}

\newcommand{\KDTree}{\mbox{$k$-d} tree\xspace}

\newcommand{\Octree}{Octree\xspace}

\newcommand{\Rtree}{\mbox{R-Tree}\xspace}

\newcommand{\Cheesemap}{\texttt{cheesemap}\xspace}

\newcommand{\chsDense}[1][]{\texttt{chs::Dense}\ifx\relax#1\relax\else\texttt{<#1>}\fi\xspace}
\newcommand{\chsSparse}[1][]{\texttt{chs::Sparse}\ifx\relax#1\relax\else\texttt{<#1>}\fi\xspace}
\newcommand{\chsMixed}[1][]{\texttt{chs::Mixed}\ifx\relax#1\relax\else\texttt{<#1>}\fi\xspace}

\newcommand{\nanoflann}{\texttt{nanoflann::KdTree}\xspace}
\newcommand{\pclkdtree}{\texttt{pcl::kdtree}\xspace}
\newcommand{\pclOctree}{\texttt{pcl::octree}\xspace}
\newcommand{\unibnOctree}{\texttt{unibn::Octree}\xspace}

\newcommand{\sdots}[1]{\kern-#1pt . \kern-#1pt . \kern-#1pt .}

\usepackage[acronym]{glossaries}
\glsdisablehyper 

\newacronym{LiDAR}{LiDAR}{Light Detection and Ranging}

\newacronym{ALS}{ALS}{Airborne Laser Scanning}
\newacronym{TLS}{TLS}{Terrestrial Laser Scanning}
\newacronym{MLS}{MLS}{Mobile Laser Scanning}
\newacronym{ULS}{ULS}{Unmanned Laser Scanning}

\newacronym{KNN}{\mbox{$k$-NN}}{$k$-nearest neighbors}

\journal{}

\begin{document}

\begin{frontmatter}



  \title{Cheesemap: A High-Performance Point-Indexing Data Structure for Neighbor~Search in LiDAR Data}


  \author[UniVie]{Ruben Laso} 
  \ead{ruben.laso.rodriguez@univie.ac.at}
  \author[inst2]{Miguel Yermo} 
  \ead{miguel.yermo@usc.es}

  \affiliation[UniVie]{organization={Research Group for Scientific Computing, Faculty of Computer Science, University of Vienna},%
    addressline={Währinger Straße 29},%
    city={Vienna},%
    postcode={1090},%
    state={Vienna},%
    country={Austria}%
  }

  \affiliation[inst2]{organization={Centro Singular de Investigación en Tecnoloxías Intelixentes (CiTIUS), Universidad de Santiago de Compostela},
    addressline={Rúa de Jenaro de la Fuente Domínguez},%
    city={Santiago de Compostela},%
    postcode={15782},%
    state={Galicia},%
    country={Spain}%
  }

  \begin{abstract}
    Point cloud data, as the representation of three-dimensional spatial information, is a fundamental piece of information in various domains where indexing and querying these point clouds efficiently is crucial for tasks such as object recognition, autonomous navigation, and environmental modeling.

    In this paper, we present a comprehensive comparative analysis of various data structures combined with neighboring search methods across different types of point clouds.
    Additionally, we introduce a novel data structure, \Cheesemap, to handle 3D LiDAR point clouds.
    Exploring the sparsity and irregularity in the distribution of points, there are three flavors of the \Cheesemap: dense, sparse, and mixed.

    Results show that the \Cheesemap can outperform state-of-the-art data structures in terms of execution time per query, particularly for ALS (Aerial Laser Scanning) point clouds.
    Memory consumption is also minimal, especially in the sparse and mixed representations, making the \Cheesemap a suitable choice for applications involving three-dimensional point clouds.
  \end{abstract}


\end{frontmatter}


\section{Introduction} \label{sec:introducion}

Point cloud data, representing three-dimensional information through discrete points, has gained significant prominence across various domains, including computer vision, robotics, archaeology, geospatial mapping, and autonomous navigation.
The efficient extraction of meaningful information from point clouds is crucial for tasks such as object recognition, scene understanding, and environmental modeling.
One of the fundamental operations in point cloud processing is the search for neighboring points around a specific point, as this operation serves as the basis for tasks like segmentation, feature extraction, data mining, convolutions, and clusterization.

The great diversity of data acquisition methods mainly using \gls{LiDAR} technology, such as \gls{ALS}, \gls{TLS}, \gls{MLS}, or \gls{ULS} has led to the creation of a wide variety of point cloud types in terms of density and distribution.
Depending on these characteristics, not all methods for neighborhood search of a given point and not all data structures used to store the point cloud will be equally suitable.

This paper presents a comprehensive comparative analysis of various data structures combined with neighboring search methods across different types of point clouds.
Our goal is to provide insights into the strengths, weaknesses, and applicability of these methods in diverse scenarios.
By understanding the performance of neighboring search techniques under various conditions, researchers and practitioners can make informed decisions when selecting an appropriate method for their specific point cloud processing tasks.
The motivation behind our research stems from the observation that state-of-the-art data structures and neighboring search methods may exhibit varying levels of efficiency depending on the nature of the point cloud data.
Factors such as varying point densities, irregular distributions, and high dimensionalities can significantly impact the performance of these methods.
Additionally, we propose a novel data structure that aims to optimize the efficiency of neighboring search operations especially (but not limited) for \gls{ALS} point clouds.

The remainder of this paper is organized as follows:
\Cref{seq:relatedwork} reviews related work on data structures and neighboring search methods in point cloud processing.
\Cref{sec:chs} introduces a novel data structure that aims to improve the search performance in \gls{ALS} point clouds.
In \cref{sec:materials}, we present the datasets used in our experiments and the experimental setup.
\Cref{sec:results} presents the results of our comparative analysis, highlighting the performance of each data structure under different conditions.
Finally, \cref{sec:conclusions} concludes the paper with a summary of key observations and directions for future work.

\section{Related work}\label{seq:relatedwork}

Computing neighborhoods in a point cloud is a non-trivial task due to the irregular nature of the data itself.
The neighborhood of a given data point is composed of all the surrounding points meeting a certain condition, and it provides information about the local structure of the point cloud, which is essential in its processing~\cite{Xiang2022}.
The most common neighboring methods encountered in the state-of-the-art literature are two:
the fixed-radius neighborhood~\cite{Turau1991} and the \gls{KNN}~\cite{Fix1989,Cover1967}.
While the former queries the point cloud to retrieve all the points inside a previously defined kernel, the latter computes the $k$-nearest points using any valid metric.

Regarding the fixed-radius neighborhoods, the typical approach is to use a spherical neighborhood, composed of all the points whose Euclidean distance to the queried point is less than $R$.
Note that different distances than the $L_2$ norm can be used, such as the $L_1$ norm or the $L_{\infty}$ norm, for generating cubic neighborhoods, for example.
Another variation is to ignore the third spatial dimension (typically the vertical, $z$), producing a cylindrical neighborhood, often used in point clouds obtained from airborne platforms~\cite{Filin2005}.
While these two kernels are the most common, any custom geometry can be defined and used to carry out the queries: cubes, and their 2D counterpart squares, toroids, and so on.

Concerning the $k$-nearest neighbors, the method needs the parameter $k$, which is the number of neighbors to be found, and the neighborhood is composed of the $k$-closest points to the queried point.
If the point being queried is located in a low-density area, the $k$-closest neighbors may include points too far away to be geometrically meaningful.
This could be detrimental to the quality of the local descriptors computed on the point, such as linearity, planarity, or eigenentropy.
To overcome this issue, some authors use a mixed approach. For example,~\cite{Wang2019} employs a fixed-radius sphere as a kernel, but only $k$ randomly selected neighbors are included.

Which type of neighborhood to use depends on the application.
When processing point clouds, the neighborhood computation is usually the most computationally expensive operation~\cite{Thomas2018}, so choosing the correct approach is crucial in terms of accuracy and saving computing time.
Both types of search methods depend on a parameter, which must be chosen carefully.
Due to the irregular nature of point clouds, a fixed-radius search can be appropriate for some areas of the cloud but may fail in others.
This occurs because of the variations of point densities across a single dataset.
To avoid this problem, some authors suggest the use of different scales across the dataset to retrieve the neighborhood.
For example,~\cite{Weinmann2015} presented a way to adapt the query parameter for each point.
An improvement of this technique is proposed in~\cite{Hackel2016, Pauly2003} for \KNN queries and in~\cite{Brodu2012, Niemeyer2014} for fixed-radius neighborhoods.
Nevertheless, the computational cost is still significant.

To reduce query times, some authors rely on stochastic sampling methods, such as the Poisson Disk Sampling~\cite{Cook1986}, which produces evenly distributed sets of points in a specific region. This is the case of~\cite{Sevgen2023}, where the random sampling is also accelerated using GPU support to keep a high number of neighbors while reducing the computation times.
To compute multiscale neighborhoods in reasonable times,~\cite{Thomas2018} opted to perform a subsampling of the dataset by using a grid, which allows them to control the size of the neighborhoods.
In the realm of real-time applications,~\cite{Tian2020} used a GPU-accelerated algorithm to detect obstacles for mobile vehicles.
The hardware accelerated methods are also useful to process massive point clouds, as is the case of~\cite{Zeng2009}, where GPUs are used to accelerate the computation of triangular irregular networks.

Notably, the method used in~\cite{Li2021} exploits the geometry of the scanlines to retrieve the spherical neighborhood in $\bigO{1}$.
By knowing the characteristics of the sensor, the point cloud is transformed into a convenient space where each point has two indexes: one for the scanline it belongs to and another for its position inside the scanline.
Nevertheless, this method is only valid when the sensor is known and its scanning pattern is regular and predictable.

\begin{figure}
  \centering
  \begin{tikzpicture}

    \definecolor{darkgray176}{RGB}{176,176,176}
    \definecolor{lightgray204}{RGB}{204,204,204}

    \begin{axis}[
        width=0.95\linewidth,
        height=5cm,
        legend cell align={left},
        legend style={
          fill opacity=1.0,
          draw opacity=1,
          text opacity=1,
          at={(0.03,0.97)},
          anchor=north west,
          draw=lightgray204
        },
        xmajorgrids,
        ymajorgrids,
        log basis x={10},
        log basis y={10},
        tick pos=both,
        x grid style={darkgray176},
        xlabel={Size of the point cloud (points)},
        xmin=0.5, xmax=2097152,
        xmode=log,
        xtick style={color=black},
        y grid style={darkgray176},
        ylabel={CPU Time (\si{\nano\second})},
        ymin=43.2831024242217, ymax=29885537.5292166,
        ymode=log,
        ytick style={color=black}
      ]
      \addplot [black]
      table {%
        1 79.7506
        8 104.771
        64 295.298
        512 1954.6
        4096 20566.1
        32768 173845
        262144 3929480
        1048576 16219800
      };
      \addlegendentry{Brute force}
      \addplot [red, dashed]
      table {%
        1 92.844
        8 106.926
        64 113.922
        512 172.339
        4096 368.943
        32768 1032.76
        262144 6066.56
        1048576 18953.4
      };
      \addlegendentry{NanoFLANN}
    \end{axis}

  \end{tikzpicture}
  \caption{Spherical query (\qty{2.5}{\meter} radius) in a synthetic point cloud of dimensions \qtyproduct{100 x 100 x 50}{\meter} on a system with a 6-core (12-threads) Intel Core i7-9750H and \qty{32}{\giga\byte} DDR4 memory at \qty{2666}{\mega\hertz}.}
  \label{fig:brute-force-vs-indexing}
\end{figure}
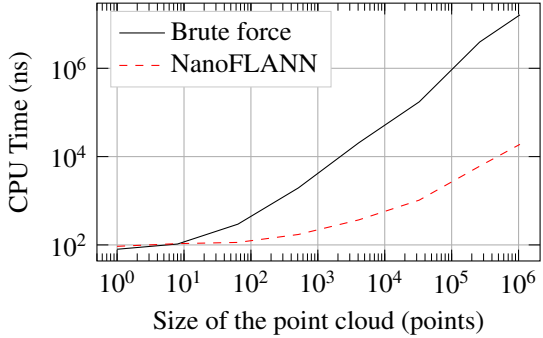

The naïve method for computing the neighborhoods is checking for every point in the dataset whether it meets the conditions to be part of the neighborhood or not.
This method could be suitable if the point clouds are composed of a few hundred points (see \cref{fig:brute-force-vs-indexing}), but modern point clouds have millions of points, making this approach unfeasible.

To overcome this issue, several data structures were created to speed up the query times.
The \KDTree~\cite{Bentley1975} is a tree-based data structure in which each node has a maximum of two children.
In the \KDTree, the data domain is partitioned by using ($k$-1)-dimensional planes, and each division created by each plane corresponds to a node of the tree.
The \Octree~\cite{Behley2015} is also a tree-based data structure, but each node has 8 children.
The space partitioning in this case is carried out using voxels: The root of the \Octree is a voxel whose size's length is the larger edge of the minimum bounding box of the dataset.
Each voxel is then subdivided into 8 equal voxels by halving its length.
The partitioning is usually stopped when a fixed number of data for a given voxel is reached.
Other not-so-common data structures are the Ball Tree~\cite{Omohundro1989} and the \Rtree~\cite{Guttman1984}.
The Ball Tree can be thought of as the same as a \KDTree, but using ($k$-1)-dimensional spheres instead of planes.
The \Rtree is similar to the \Octree, but the space partitioning is carried out by computing smaller bounding boxes for each level of the tree.
To check whether two elements are nearby, the collision of the bounding boxes must be computed, which usually is a computationally cheap operation.

The most commonly used of the aforementioned data structures to carry out neighboring searches are the \KDTree and the \Octree.

\KDTree is known for its effectiveness in multidimensional data for nearest-neighbor searches~\cite{Li2021}, having an average complexity of $\bigO{\log{N}}$ for finding the nearest neighbor in a point cloud with $N$ points.
Nevertheless, this complexity can be limiting~\cite{Luo2022}.

\Octree is widely utilized for organizing point clouds into voxels, simplifying the exploitation of its structure for neighbor searching~\cite{Che2019}.
It is also used in indexing structures for massive point clouds, enhancing the efficiency of spatial indexing~\cite{Wang2021}, and neighbor searching methods for 3D point clouds, particularly in the context of \gls{LiDAR} data~\cite{Feng2013}.

Furthermore, in the context of graph-based networks and point cloud analysis, \KDTree is utilized for conducting point-to-node $k$-nearest-neighbor searches.
This allows for the systematic adjustment of the network's receptive field~\cite{Li2018}.
Additionally, \KDTree is used in the establishment of a three-dimensional point cloud registration based on entropy and particle swarm optimization, where the relationship of points is established by \KDTree for finding the $k$-nearest neighbor of a point~\cite{Zhan2018}.
\Octree, on the other hand, has been used in the context of 3D change detection, where it is employed for adaptive thresholds based on local point cloud density~\cite{Liu2021}.

Both \KDTree and \Octree play crucial roles in neighbor searches in point clouds, however, to the best of our knowledge, they have been reviewed only twice in the literature.
In the work by Elseberg et al.~\cite{Elseberg2012}, the performance of five implementations of \KDTree, one \Octree, and one \Rtree is compared when used for k-nearest neighbors and fixed-radius queries, in the context of shape registration in synthetic and real point cloud data.
The other review is presented by Lawson et al.~\cite{Lawson2022}, where a single unstructured mesh composed of \num{152.7} million nodes and \num{152.1} million hexahedral elements is used as a study case.
In that review, a deep comparison between 20 open-source C/\Cpp libraries implementing different search methods is presented.
However, the study uses synthetic data with an unknown spatial distribution.

While the aforementioned studies presented are exhaustive, some questions remain unanswered:
How do the different data structures perform in real-world point clouds?
How do they perform in different types of point clouds?
How do the density and distribution of points affect the performance of neighboring search methods?
Is the performance different in \KNN and fixed-radius queries?
Can a novel data structure outperform current state-of-the-art data structures?

In this work, we aim to answer these questions by comparing different methods for performing neighborhood search, while proposing a new data structure to perform neighbor searches in three-dimensional point clouds.

\section{Cheesemap}\label{sec:chs}

As an alternative to the proposals above, we present a data structure known as \Cheesemap and its flavors: dense, sparse, and mixed.

This structure, particularly in its sparse and mixed forms, aims to take advantage of the features of \gls{LiDAR} point clouds, in which point density is usually not uniform.
In \gls{ALS} point clouds, the range in the vertical $z$-axis tends to be significantly smaller than in the $x$- and $y$-axis.
Thus, when indexing the points, the $z$-axis can be omitted.
In other point clouds, like \gls{TLS} or \gls{MLS}, and particularly in urban environments, points do not follow a uniform distribution, with the density varying significantly depending on the distance to the sensor and the presence of obstacles.

Note that the \Cheesemap works aligned with the $x$, $y$, and $z$ axes.
If the point cloud is not aligned with these axes, a rotation could be desirable for using the \Cheesemap efficiently.
This is a common limitation in other data structures, such as the \KDTree and the \Octree, which are also axis-aligned.

\subsection{Notation convention} \label{sec:chs-notation}
In the rest of the paper, we use the $x$, $y$, and $z$ subscripts to denote the dimension.
For example, $p_x$ would reference the $x$ coordinate of point $p$.

Additionally, we define $p^{-}$ and $p^{+}$ as the corners of the bounding box of the point cloud $P$, such as:
\begin{align}
  p^{-} & = \tuple{\min\set{p_a \mid p \in P}, \ a = \set{x, y, z}},
  \\
  p^{+} & = \tuple{\max\set{p_a \mid p \in P}, \ a = \set{x, y, z}}.
\end{align}

\subsection{General idea} \label{sec:chs-idea}

The basic structure of the \Cheesemap relies on 2D and 3D voxels, depending on the coordinates used to index.

For the sake of conciseness, explanations in this paper will assume three-axis coordinates.
The same calculations are also valid for 2D by ignoring the third axis.
Thus, if only $x$ and $y$ coordinates are used for indexing the points, 2D voxels are used.
If $x$, $y$, and $z$ coordinates are being used, the \Cheesemap uses 3D voxels.

Let $s$ be the size of the voxel\footnote{In the explanation, we assume that the dimensions of the voxel are the same in all directions for the sake of simplicity. In the actual implementation, they may vary.} and let $B = \tuple{p^{-}, p^{+}}$ the bounding box of the point cloud.
Then, a grid $G$ of $\tuple{n_{x}, n_{y}, n_{z}}$ voxels is generated such that,
\begin{gather}
  n_{a} = \floor{\left(p_{a}^{+} - p_{a}^{-}\right) / s} + 1, \ a = \set{x, y, z}, \label{eq:grid-size}
  \\
  G := \set{v_{ijk},\ i = 1, \sdots{0.25}, n_{x},\ j = 1, \sdots{0.25}, n_{y},\ k = 1, \sdots{0.25}, n_{k}}. \label{eq:grid}
\end{gather}

As a consequence of the structure of the grid, given a point $p$, the indices of its correspondent voxel $v_{ijk}$ are determined by
\begin{gather}
  \set{i, j, k} = \set{\floor{\left(p_{a} - p_{a}^{-}\right) / s}, \ a = \set{x, y, z}}. \label{eq:chs-point-to-idcs}
\end{gather}

\subsection{Flavors of \Cheesemap} \label{sec:chs-flavours}

The most interesting part of the \Cheesemap lies in the way of storing the voxels and their information.
To do so, three variants are proposed: dense, sparse, and mixed.

\subsubsection{Dense} \label{sec:chs-dense}
The dense representation of \Cheesemap is the most traditional one, where a list of $n_x \times n_y \times n_z$ voxels is built.
Storing these voxels is straightforward thanks to the regular structure of the grid.
Thus, it is possible to identify voxels by a single global index,
\begin{gather}
  g(i, j) = i n_y + j, \label{eq:indices-to-global-2D}
  \\
  g(i, j, k) = i (n_y n_z) + j (n_z) + k, \label{eq:indices-to-global-3D}
\end{gather}
for the two- and three-dimensional cases, respectively.
These global indices are used to know the position of a voxel in an array for its storage and lookup.

\subsubsection{Sparse} \label{sec:chs-sparse}
The sparse representation stores only non-empty voxels (those containing at least one point).

\begin{figure}
  \centering
  \begin{tikzpicture}[scale=0.95]
    \foreach \i in {1,...,1000}
    \fill[black!80] (rnd*7.90+0.05,rnd*4.90+0.05) circle (0.05);

    \draw[fill=gray!30,fill opacity=1.0] (4.3,2.4) ellipse (2 and 1.5);

    \foreach \x in {0,1,...,7}
    \foreach \y in {0,1,...,4}
    \draw[black!60] (\x,\y) rectangle (\x+1,\y+1);

    \draw[fill=gray!30, pattern=north west lines, pattern color=black] (3,2) rectangle (4,3);
    \draw[fill=gray!30, pattern=north west lines, pattern color=black] (4,2) rectangle (5,3);
    \draw[fill=gray!30, pattern=north west lines, pattern color=black] (5,2) rectangle (6,3);
    \draw[fill=gray!30, pattern=north west lines, pattern color=black] (4,1) rectangle (5,2);

  \end{tikzpicture}
  \caption{Example of sparsity in a point cloud. There are no points within the light-gray ellipse (representing, for example, a lake in ALS point clouds), so shaded voxels do not contain any points.}
  \label{fig:chs-sparse-example}
\end{figure}
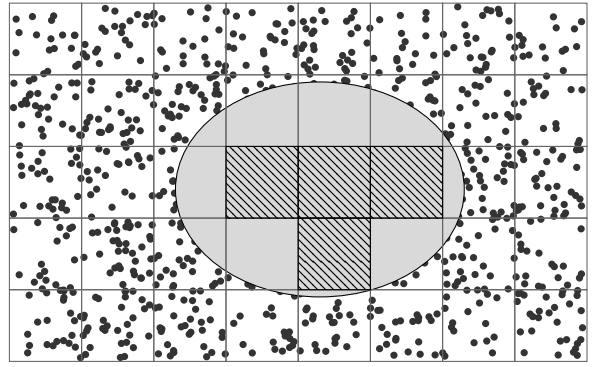

Imagine, for example, a point cloud like the one shown in \cref{fig:chs-sparse-example} where there is a big empty region.
Those voxels fully contained within that region do not need to be stored as they contain no points.

To store only those non-empty voxels, a hashmap is used, where the index of the map corresponds to the global index given by \cref{eq:indices-to-global-2D,eq:indices-to-global-3D}.
The reason behind that choice is that a hashmap provides amortized constant time for lookup and accessing its contents.

\subsubsection{Mixed} \label{sec:chs-mixed}
The mixed structure combines the dense and sparse representation by dividing the point cloud into slices that can be either dense or sparse.

The point cloud is divided into $n_k$ slices through the $z$-axis.
Since the $z$-axis is typically the smallest dimension, fewer slices are generated by splitting the point cloud in this direction, speeding up the lookup process.
Note that a smaller $s_{z}$ can be used to generate more slices if needed.

Another advantage of slicing along the $z$-axis is that, in \gls{LiDAR} point clouds, horizontal planes near the ground typically have a higher density than those at higher elevations.
This way, those slices that are closer to the ground could take advantage of a faster lookup with the dense representation, while those above the ground would use a more memory-efficient indexing method with sparse indexing.

At first, all slices are represented in sparse form, in the same way as shown in \cref{sec:chs-sparse}.
As points are added to the voxels of each slice, the number of non-empty voxels is counted.
Once a slice has enough non-empty voxels, it is turned into a dense slice.
By default, this conversion happens when \qty{80}{\percent} of the voxels have points.
This optimizes memory usage for low-density slices, and lookup speed for highly-populated slices.

In the 2D mixed \Cheesemap, a single slice is generated ($n_k=1$).
This slice follows the same rules as in the 3D case: once it has sufficient non-empty voxels, it is changed from the sparse to the dense representation.

\subsection{Neighborhood search} \label{sec:chs-search}
Thanks to the structured grid of the \Cheesemap, queries can be performed efficiently, as the time complexity to find the voxel containing a point is $\bigO{1}$ for the dense \Cheesemap, and amortized $\bigO{1}$ for the sparse and mixed versions.
In this section, we delve into the implementation of the two most common operations: kernel-based and $k$-nearest neighbors search.

\subsubsection{Kernel-based search} \label{sec:chs-kernel-search}
Imagine a spherical kernel\footnote{The same would apply to other kernels like boxes.} defined by the ball $B(c, r)$, centered at $c$ with radius $r$.
The first step is to compute the bounding box of the kernel, $\tuple{p^-_\text{ker}, p^+_\text{ker}}$.
In our example:
\begin{gather}
  p^-_\text{ker} = \tuple{c_x - r, c_y - r, c_z - r}
  \\
  p^+_\text{ker} = \tuple{c_x + r, c_y + r, c_z + r}
\end{gather}

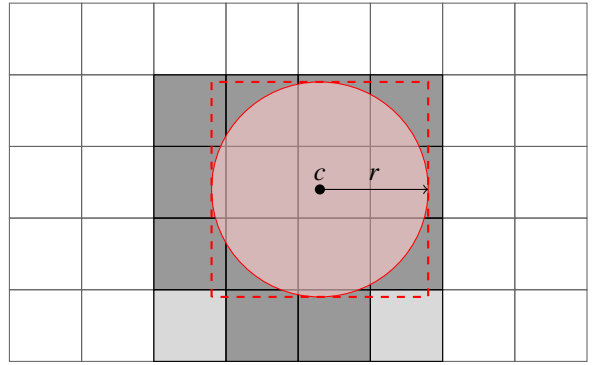
\begin{figure}[bt]
  \centering
  \begin{tikzpicture}[scale=0.95]
    \foreach \x in {0,1,...,7}
    \foreach \y in {0,1,...,4}
    \draw[black!60] (\x,\y) rectangle (\x+1,\y+1);

    \tikzstyle{square} = [draw, fill opacity=0.5, minimum width=1cm, minimum height=1cm]
    \tikzstyle{dsquare} = [square, fill=black!80]
    \tikzstyle{lsquare} = [square, fill=black!30]

    \draw[lsquare] (2,0) rectangle (3,1);
    \draw[dsquare] (2,1) rectangle (3,2);
    \draw[dsquare] (2,2) rectangle (3,3);
    \draw[dsquare] (2,3) rectangle (3,4);

    \draw[dsquare] (3,0) rectangle (4,1);
    \draw[dsquare] (3,1) rectangle (4,2);
    \draw[dsquare] (3,2) rectangle (4,3);
    \draw[dsquare] (3,3) rectangle (4,4);

    \draw[dsquare] (4,0) rectangle (5,1);
    \draw[dsquare] (4,1) rectangle (5,2);
    \draw[dsquare] (4,2) rectangle (5,3);
    \draw[dsquare] (4,3) rectangle (5,4);

    \draw[lsquare] (5,0) rectangle (6,1);
    \draw[dsquare] (5,1) rectangle (6,2);
    \draw[dsquare] (5,2) rectangle (6,3);
    \draw[dsquare] (5,3) rectangle (6,4);

    \draw[red, fill=red!20, fill opacity=0.6] (4.3,2.4) circle (1.5);
    \fill (4.3,2.4) circle (2pt) node[anchor=south] {$c$};
    \draw[->] (4.3,2.4) -- (5.8,2.4) node[midway,above] {$r$};

    \draw[dashed, red, thick] (2.8,0.9) rectangle (5.8,3.9);

  \end{tikzpicture}
  \caption{Example of kernel-based search. The red circle represents the kernel, a sphere centered in $c$ with radius $r$.
    The red dashed box represents the bounding box of the sphere.
    In light gray, the voxels that intersect with the bounding box, and may intersect with the kernel.
  In dark gray, those voxels that actually intersect with the sphere.}
  \label{fig:chs-kernel-search-example}
\end{figure}

With the bounding box, it is possible to determine which voxels should be explored (as they may intersect with the kernel) using \cref{eq:chs-point-to-idcs}.
An example of the intersection region is shown in \cref{fig:chs-kernel-search-example}.

After calculating the intersection region, it is only a matter of traversing through the points in those voxels, and returning the list of points within the search kernel.

Note that the same procedure can be applied to any kernel shape, as long as the bounding box is provided.
In the code, the kernel must comply with a \Cpp20 concept that requires the definition of the functions \verb|box()|, which returns the bounding box of the kernel, and \verb|is_inside()|, which returns whether a point is inside the kernel.

\Cref{alg:chs-kernel-search} provides the pseudo-code of the kernel-based search.

\begin{algorithm}
  \caption{Kernel-based neighborhood search} \label{alg:chs-kernel-search}
  \begin{algorithmic}[1] 
    \Require Function $\texttt{is\_inside}(p)$ returning true is $p$ is inside the kernel.
    \State $P_\text{neigh} = \varnothing$ \Comment{List of neighbors.}
    \State $\tuple{p^{-}_\text{ker}, p^{+}_\text{ker}} =$ kernel bounding box
    \State $\tuple{i^{-}, j^{-}, k^{-}} =$ indices such $p^{-}_\text{ker} \in   v_{ijk}$
    \State $\tuple{i^{+}, j^{+}, k^{+}} =$ indices such $p^{+}_\text{ker} \in  v_{ijk}$
    \For{$\tuple{i, j, k} = \tuple{i^{-}, \dots, i^{+}} \times \tuple{j^{-}, \dots, j^{+}} \times \tuple{k^{-}, \dots, k^{+}}$}
    \If{$\exists v_{ijk}$} \label{line:voxel-exists-kernel-search}
    \ForAll{$p \in v_{ijk}$}
    \If{$\texttt{is\_inside}(p)$}
    \State $P_\text{neigh} := P_\text{neigh} \cup \set{p}$
    \EndIf
    \EndFor
    \EndIf
    \EndFor{}
    \State \Return $P_\text{neigh}$
  \end{algorithmic}
\end{algorithm}

\subsubsection{\KNN search} \label{sec:chs-knn-search}
When looking for the \KNN, two kinds of points are considered:
\begin{itemize}
  \item Candidate points: points whose distance to the center point has already been calculated.
  \item Non-visited points: points whose distance has not yet been computed.
\end{itemize}
Thus, a set $P_\text{cand}$ is used for the candidate points, and a subset of it, $\hat{P}_\text{cand}$, which stores the points that are within a distance $r$ from the center point $c$:
\begin{equation} \label{eq:chs-knn-cand-within-radius}
  \hat{P}_\text{cand} = \set{p \in P_\text{cand} \mid d(p, c) \leq r}.
\end{equation}
Additionally, a set of voxels named $T$ is defined and serves as a taboo list of already visited voxels to avoid redundant calculations.

To find the $k$-nearest neighbors of a point $c$, a two-phase iterative search is performed where each iteration is similar to a spherical kernel-based search with an increasing search radius $r$.
In each iteration, the voxels that intersect with the search region $B(c, r)$ are visited, the points contained in those voxels are added to $P_\text{cand}$, and $r$ is updated.
Because of the taboo list $T$, each voxel is visited only once, so the distances of its points to $c$ are calculated a single time.
The search stops when the condition $\lvert \hat{P}_\text{cand} \rvert \geq k$ is met.

In the first phase, a monotonic growing search is performed to find the initial neighbors.
The starting radius $r$ corresponds to the closest distance from $c$ to the wall of its voxel:
\begin{gather}
  r = \min\set{\abs{c_a - p^{-}_{v,a}}, \abs{c_a - p^{+}_{v,a}}, \ a = \set{x, y, z}},
\end{gather}
where $p^{+}_{v}$ and $p^{-}_{v}$ are the delimiting points of the voxel.
After the initial iteration, the radius is continuously increased by the size of the voxel $s$ until the search region is not empty and $\hat{P}_\text{cand} \neq \varnothing$.

In the second phase, the density of $\hat{P}_\text{cand}$ is used to estimate how large $r$ should be to find $k$ neighbors:
\begin{gather}
  \rho = \frac{|\hat{P}_\text{cand}|}{(4/3) \pi r^3},
  \\
  r = \left(\frac{\rho}{k}\right)^{1/3}.
\end{gather}
If the density does not change between two consecutive iterations, the radius is updated by $r = r + s$.
This second phase is repeated until we find $k$-nearest neighbors, $\lvert \hat{P}_\text{cand} \rvert = k$, or all the voxels have been visited, $T = G$.

\Cref{alg:chs-knn-search} shows the pseudocode of the \KNN search.

\begin{algorithm*}[t]
  \caption{\KNN search} \label{alg:chs-knn-search}
  \begin{algorithmic}[1] 
    \Require Number of neighbors $k$.
    \Require Point $c$ used as center of the \KNN neighborhood.
    \Require Function $d(p, q)$ that returns the distance between points $p$ and $q$.
    \State $P_\text{cand} = \varnothing$ \Comment{List of points whose distance to $c$ has been calculated}
    \State $\hat{P}_\text{cand} = \varnothing$ \Comment{Points of $P_\text{cand}$ within the search radius}
    \State $T = \varnothing$ \Comment{Taboo list for the already computed voxels}
    \State $r = \min\set{\abs{c_a - p^{-}_{v,a}}, \abs{c_a - p^{+}_{v,a}}, \ a = \set{x, y, z}}$ \Comment{Initial radius}
    \While{$|\hat{P}_\text{cand}| < k$ and $T \neq G$} \Comment{$\hat{P}_\text{cand}$ defined in \cref{eq:chs-knn-cand-within-radius}}
      \State $\tuple{p^{-}_\text{ker}, p^{+}_\text{ker}} =$ bounding box of $B(c, r)$
      \State $\tuple{i^{-}, j^{-}, k^{-}} =$ indices such $p^{-}_\text{ker} \in  v_{ijk}$
      \State $\tuple{i^{+}, j^{+}, k^{+}} =$ indices such $p^{+}_\text{ker} \in  v_{ijk}$
      \For{$\tuple{i, j, k} = \tuple{i^{-}, \dots, i^{+}} \times \tuple{j^{-}, \dots, j^{+}} \times \tuple{k^{-}, \dots, k^{+}}$}
        \If{$\exists v_{ijk}$ and $v_{ijk} \notin T$} \label{line:voxel-exists-knn-search}
          \State $T = T \cup \set{v_{ijk}}$
          \ForAll{$p \in v_{ijk}$}
            \State $P_\text{cand} := P_\text{cand} \cup \set{p}$
          \EndFor
        \EndIf
      \EndFor
      \State $\hat{P}_\text{cand} = \set{p \in P_\text{cand} \mid d(p, c) \leq r}$ \Comment{Update $\hat{P}_\text{cand}$}
      \If{$\hat{P}_\text{cand} \neq \varnothing$}
        \State $\rho = \frac{|\hat{P}_\text{cand}|}{(4/3) \pi r^3}$ \Comment{Density-based growth}
        \State $r = \left(k / \rho \right)^{1/3}$
      \Else
        \State $r = r + s$ \Comment{Monotonic growth}
      \EndIf
    \EndWhile
    \State \Return $\hat{P}_\text{cand}$
  \end{algorithmic}
\end{algorithm*}

\subsection{Further remarks} \label{sec:chs-optimizations}
As part of the design and implementation of the \Cheesemap, some details have been considered to enhance its performance.

\paragraph{Reordering}
To enhance the cache-friendliness of the \Cheesemap, the points of the dataset can be reordered according to the voxel they belong to (see \cref{eq:indices-to-global-2D,eq:indices-to-global-3D}).
Then, when iterating over the points of a voxel, their memory addresses should be closer or even consecutive, improving the cache hit rate and theoretically speeding up the search.

\paragraph{Voxels in dense \Cheesemap}
It is worth noting that all voxels are created in the dense version of \Cheesemap, so it is not necessary to check if a certain voxel exists (Lines \ref{line:voxel-exists-kernel-search} and \ref{line:voxel-exists-knn-search} of \cref{alg:chs-kernel-search,alg:chs-knn-search}, respectively).

\paragraph{Hashmap vs sparse matrix in sparse \Cheesemap}
Initially, a sparse matrix was considered to store the voxels in the sparse version of \Cheesemap.
However, a hashmap was preferred due to its better performance in amortized constant time for lookup and insertion.
In its current implementation, \Cheesemap uses the \verb|std::unordered_map| from the \Cpp Standard Library.

\paragraph{Taboo list in \KNN}
As mentioned in \cref{sec:chs-kernel-search}, the taboo list not only serves to obtain correct results (avoiding duplications) but also affects performance by preventing the repetition of calculations.
Thanks to the structured grid, the implementation of the taboo list can be optimized by storing only the indices $\tuple{i^{-}, j^{-}, k^{-}}$ and $\tuple{i^{+}, j^{+}, k^{+}}$ of the voxels that have been visited.
In the next iteration, only the new indices that do not belong to the last $(i^{-}, i^{+}) \times (j^{-}, j^{+}) \times (k^{-}, k^{+})$ region need to be visited.
This optimization reduces both memory consumption and execution time, as only two tuples of indices are stored instead of a (potentially large) list of voxels, and the comparison of the indices is faster than the lookup in a list.

\paragraph{Candidate list in \KNN}
In the current implementation, both sets $P_\text{cand}$ and $\hat{P}_\text{cand}$ are stored in the same list, using a custom data structure built on top of a \verb|std::vector|.
The container is designed to store pairs of points and distances, $\tuple{p, d(p, c)}$, and its elements are sorted by distance.
When inserting a new point, a binary search is performed to find the correct position in the list.
Then, the point is inserted and the elements with a greater distance are shifted one position to the right.
When the list is full (it has more than $k$ elements), the last element is removed.
Despite the worst-case complexity of this operation being $O(n)$, this is not the case in reality.
The reason is that, with an increasing-radius search, the new points to be inserted are further and further away from the center, so their position in the list is closer to the end, and the number of elements to be shifted is small (if any at all).

\paragraph{Density-based growth in \KNN}
Experimentally, it has been observed that the growth of the search radius using the density of the neighbors is more efficient than monotonic growth.
In a synthetic dataset of \num{1} million random points, the search radius increased by density reduced the number of explored voxels in \qty{84.5}{\percent} of the cases, while a monotonic growth was better in only \qty{3.6}{\percent} of the cases.
In the remaining \qty{11.9}{\percent} of the cases, both methods were equivalent.
Consequently, the algorithm is slightly faster when using density-based growth.


\section{Materials and Methods}\label{sec:materials}

In this section, we describe the datasets used, the machines where the experiments will be conducted, and the metrics used to evaluate the performance of the data structures.

\subsection{Datasets}\label{subsec:datasets}

We chose a set of point clouds representative of real-world applications to compare various state-of-the-art data structures and their implementations by different authors.
Our goal was to choose datasets that are open, publicly available, and widely used in the literature.
For this purpose, we have selected several datasets representing different environments for each type of platform.
The density of each dataset is computed as the ratio of the total number of points to the area of the bounding box.
For the weighted density, we calculate the local density of each point as the ratio of its number of neighbors to the area of its neighborhood.
In this study, a 2D circular neighborhood with a radius of 1 meter is used.
A histogram with 256 bins is then created, with each bin representing an integer value of density.
The weighted density is obtained by averaging the local densities, where each local density is weighted according to its frequency in the histogram.
\Cref{tab:datasets} shows a summary of the dataset used in this study.

\begin{table*}[tb]
  \centering
  \caption{Characteristics of the datasets used in the comparison. The sensor information used in the Semantic3D dataset is not publicly available.}
  \label{tab:datasets}
  \begin{adjustbox}{max width=\linewidth}
    \begin{tabular}{lccS[table-format=9.0]ccS[table-format=5.2]S[table-format=3.2]}
      \toprule
      Name         & Type & Sensor            & {Points}  & Extension                         & Scene    & {Density}                       & {Weighted density}              \\
      &      &                   &           &                                   &          & {(\si{\points/\meter\squared})} & {(\si{\points/\meter\squared})} \\
      \midrule
      \dalesA      & ALS  & Riegl Q1560       & 12219779  & \qty{0.250}{\kilo\meter\squared}  & Urban    & 48.87                           & 61.00                           \\
      \dalesB      & ALS  & Riegl Q1560       & 17747769  & \qty{0.250}{\kilo\meter\squared}  & Urban    & 70.99                           & 69.93                           \\
      \dalesC      & ALS  & Riegl Q1560       & 11981458  & \qty{0.250}{\kilo\meter\squared}  & Rural    & 47.93                           & 20.68                          \\
      \dalesD      & ALS  & Riegl Q1560       & 12148800  & \qty{0.250}{\kilo\meter\squared}  & Rural    & 48.60                           & 57.90                           \\
      \Lille       & MLS  & Velodyne HDL-32E  & 80000000  & \qty{0.798}{\kilo\meter\squared}  & Urban    & 100.27                          & 49.13                           \\
      \Paris       & MLS  & Velodyne HDL-32E  & 50000000  & \qty{0.048}{\kilo\meter\squared}  & Urban    & 1030.83                         & 42.78                          \\
      \Speulderbos & ULS  & Riegl VUX-1UAV    & 79221314  & \qty{0.033}{\kilo\meter\squared}  & Forestry & 2378.46                         & 109.80                           \\
      \Hallway     & TLS  & Matterport Camera & 3840664   & \qty{151.089}{\meter\squared}     & Indoor   & 25419.88                        & 120.48                           \\
      \sgRural     & TLS  & ---               & 170158281 & \qty{0.262}{\kilo\meter\squared}  & Rural    & 649.85                          & 101.67                           \\
      \sgUrban     & TLS  & ---               & 258720948 & \qty{0.050}{\kilo\meter\squared}  & Urban    & 5181.00                         & 73.24                           \\
      \bottomrule
    \end{tabular}
  \end{adjustbox}
\end{table*}

The DALES~\cite{Varney2020} dataset is a well-known benchmark for evaluating deep learning classification models acquired using airborne laser scanning.
The dataset is split into squared chunks of \qty{250000}{\meter^2}.
For the purposes of this work, four chunks were selected representing different types of areas: a residential area with low buildings (\dalesA), a downtown area with skyscrapers (\dalesB), an isolated area with high-voltage powerlines (\dalesC), and a rural area with high vegetation (\dalesD).

Mobile laser scanning is often used to capture urban environments, which are well represented by the Paris-Lille-3D~\cite{Roynard2018} dataset, from which we used the scenes of \Paris and \LilleTwo (referred to as \Lille for simplicity).

To the best of our knowledge, there are no reference benchmark datasets for unmanned laser scanning.
We have chosen the \Speulderbos~\cite{Brede2019} dataset, which contains around \qty{80}{\mega\SIvoid} points, despite covering a relatively small area (\qty{0.018}{\kilo\meter\squared}).
Its utility lies in analyzing the behavior of data structures in extremely dense point clouds.

Regarding terrestrial laser scanning, two datasets were chosen.
As an example of an indoor point cloud taken from the S3DIS~\cite{Armeni2016} dataset, we have chosen the one containing the largest number of \gls{LiDAR} points: \Hallway.
For outdoor areas, two scenes from the Semantic3D~\cite{Hackel2017} dataset were selected.
The \sgRural scene corresponds to a farm in a rural area, and \sgUrban is a point cloud representing an urban environment.
Other scenes in this dataset correspond to streets, which are better represented in mobile laser scanning datasets.
The sensor information used for the acquisition of this dataset was not publicly disclosed.

The datasets are distributed in LAS, ASCII, and binary PLY formats. All were converted to the standard LAS format.

\subsection{Tested libraries}
To assess the performance of the \Cheesemap, we have compared it against several state-of-the-art libraries and data structures.
These libraries were chosen based on their popularity and the availability of their source code.
Below, we present the libraries used in the comparison:
\begin{itemize}
  \item The \verb|nanoflann| library~\cite{blanco2014nanoflann} is a \Cpp header-only library based on the original FLANN~\cite{DBLP:conf/visapp/MujaL09}, optimizing the performance of queries using a \KDTree. In the results, we will refer to this library as \nanoflann.
  \item The \verb|PCL| library~\cite{PCL2011} is a widely-used library for point cloud processing.
    It includes several data structures for point cloud indexing, such as the \KDTree and the \Octree, which will be noted as \pclkdtree and \pclOctree, respectively.
  \item The implementation of the \Octree by Behley \textit{et al.} at the University of Bonn~\cite{Behley2015}, which is a reference in the field of point cloud processing. This implementation will be referred to as \unibnOctree.
  \item The implementations of the \Cheesemap, which will be referred to as \chsDense, \chsSparse, and \chsMixed. Dimensionality is noted within angle brackets and if the point cloud is reordered it will be indicated with a suffix. For example, ``\chsDense[3] (reordered)'' refers to the 3D dense version with reordering.
\end{itemize}

\subsection{Experimental environment}\label{subsec:machines}
The experiments were conducted on a NUMA system with four Intel Xeon E5-4620 v4 processors at \qty{2.1}{\giga\hertz} and 10 physical cores each.
Hyperthreading was disabled during the experiments.
The total installed memory was \qty{256}{\gibi\byte} DDR4.
The operating system used was AlmaLinux 8.6 with kernel~4.18.0.

\subsection{Methods}
To compare the performance of the different data structures, we conducted several experiments measuring the time needed to perform a spherical-kernel search, a cubic-kernel search, and a \KNN search.
To that end, we considered the following radii: \qtylist{0.5;1.0;2.0;3.0;5.0;7.5;10.0}{\meter}, and the following number of neighbors: \numlist{5;10;20;30;40;50}.

Each query consisted of selecting a random point from the point cloud and performing the search with the specified radius or number of neighbors.

Using Google's benchmark library\footnote{Available at \url{https://github.com/google/benchmark}.}, each experiment ran for \qty{1}{\second}, during which as many queries as possible were performed.
For example, if a query takes on average \qty{0.1}{\milli\second}, approximately \num{10000} different queries will be performed.
The limit of \qty{1}{\second} was deemed sufficient to obtain reliable average times, since each query is expected to take between \si{\micro\second} and \si{\milli\second}.

The memory footprint of the different data structures was also measured using Malt~\cite{Malt2017}.

As mentioned in \cref{sec:chs}, some data structures (including our proposal) could benefit from applying certain rotations or transformations to the point cloud.
However, we decided to avoid these optimizations to ensure a fair comparison between the different structures.

\section{Results and discussion}\label{sec:results}
The results of the experiments are divided into two sections, one analyzing the performance of the \Cheesemap and its parameters, and another comparing the \Cheesemap with other state-of-the-art data structures.

\subsection{Performance of \Cheesemap} \label{sec:cheesemap-performance}
First, we analyze how the performance of the dense, sparse, and mixed implementations of the \Cheesemap varies across different datasets and parameters.

\begin{figure*}
  \centering
  \includegraphics[width=\linewidth]{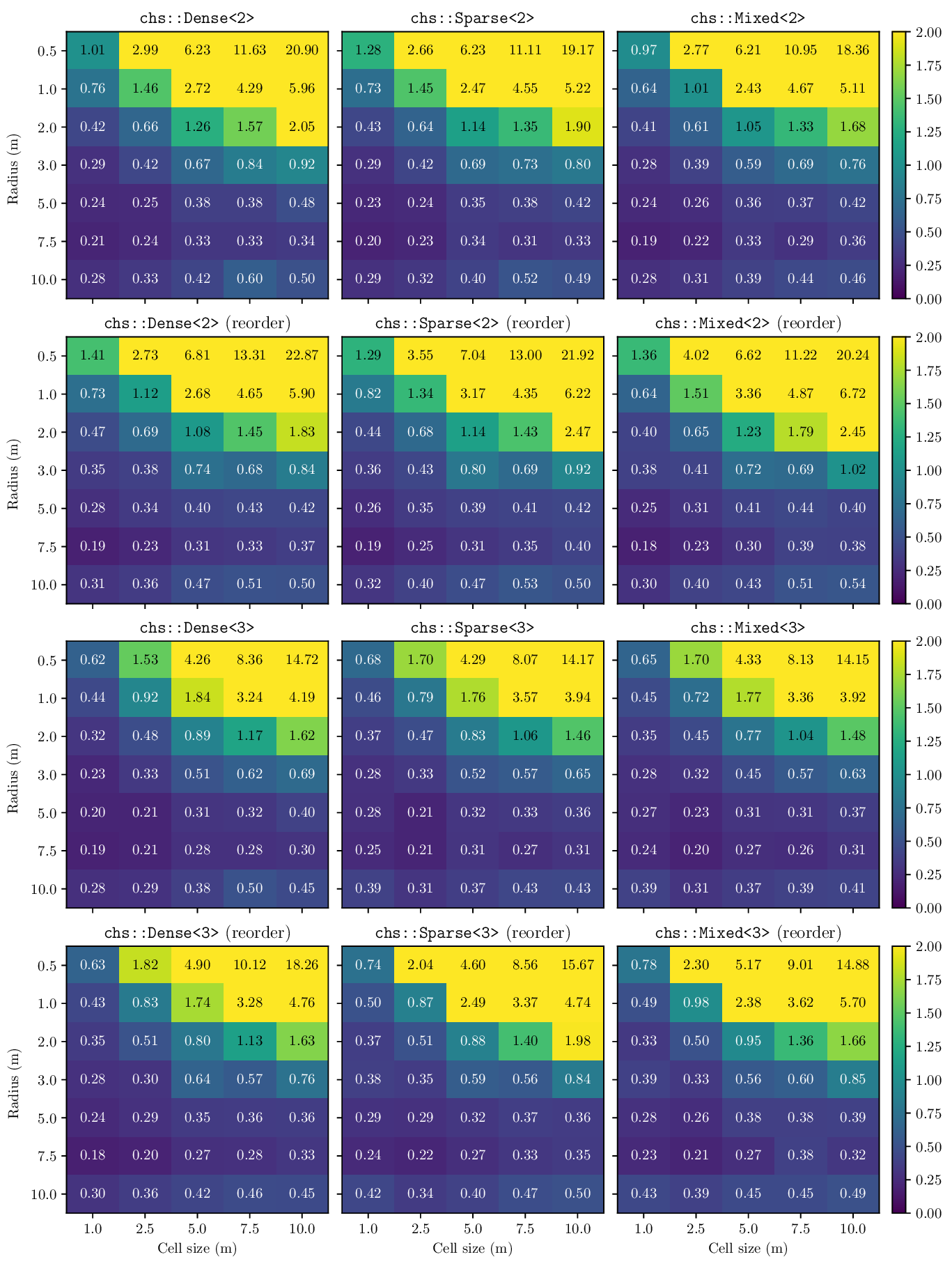}
  \caption{Geometric average of the normalized search time of the \Cheesemap against \nanoflann in the spherical-kernel search. Lower is better.}
  \label{fig:geom-avg-vs-nanoflann-sphere}
\end{figure*}

\begin{figure*}
  \centering
  \includegraphics[width=\linewidth]{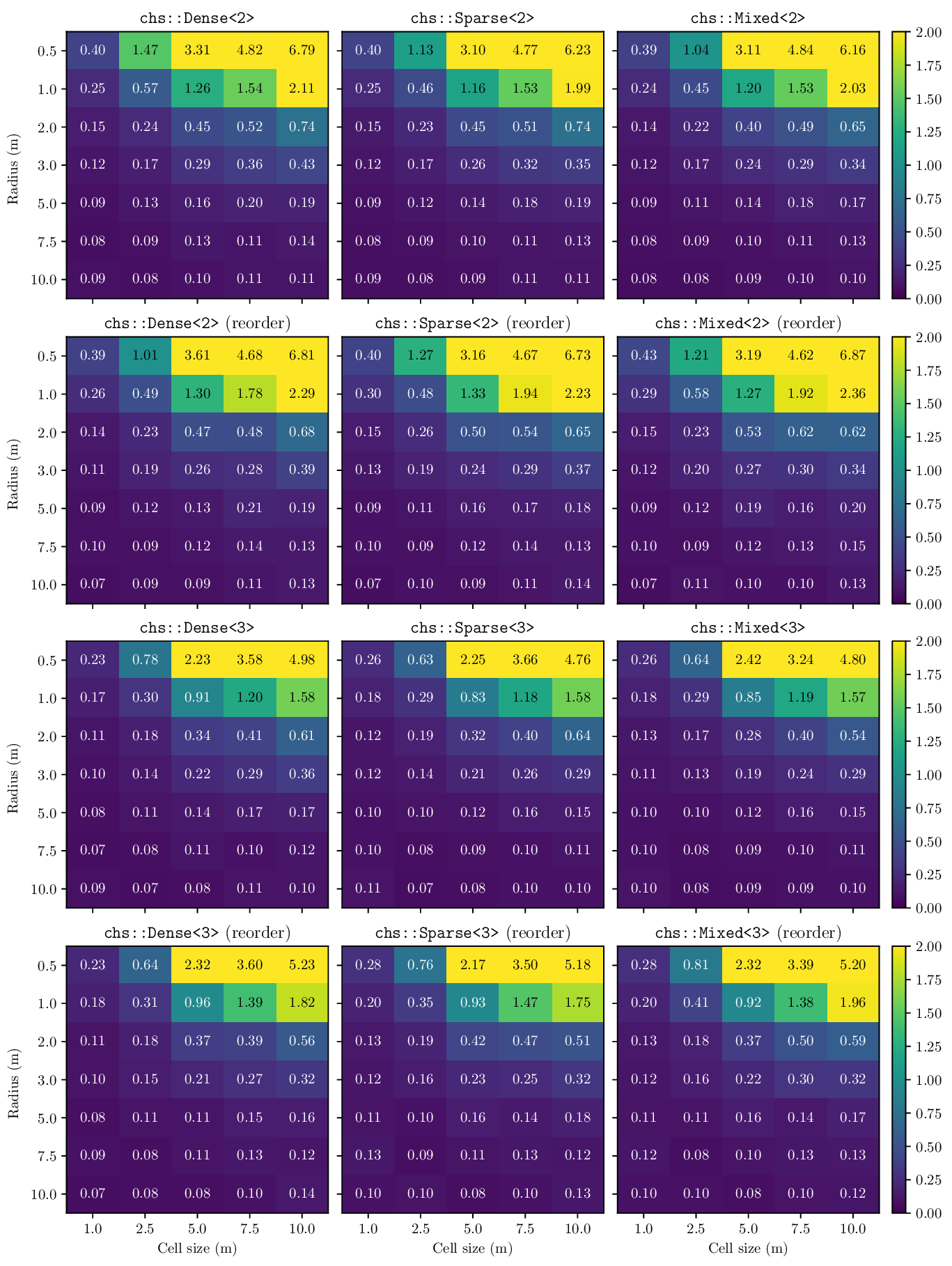}
  \caption{Geometric average of the normalized search time of the \Cheesemap against \nanoflann in the cube-kernel search. Lower is better.}
  \label{fig:geom-avg-vs-nanoflann-cube}
\end{figure*}

\begin{figure*}
  \centering
  \includegraphics[width=\linewidth]{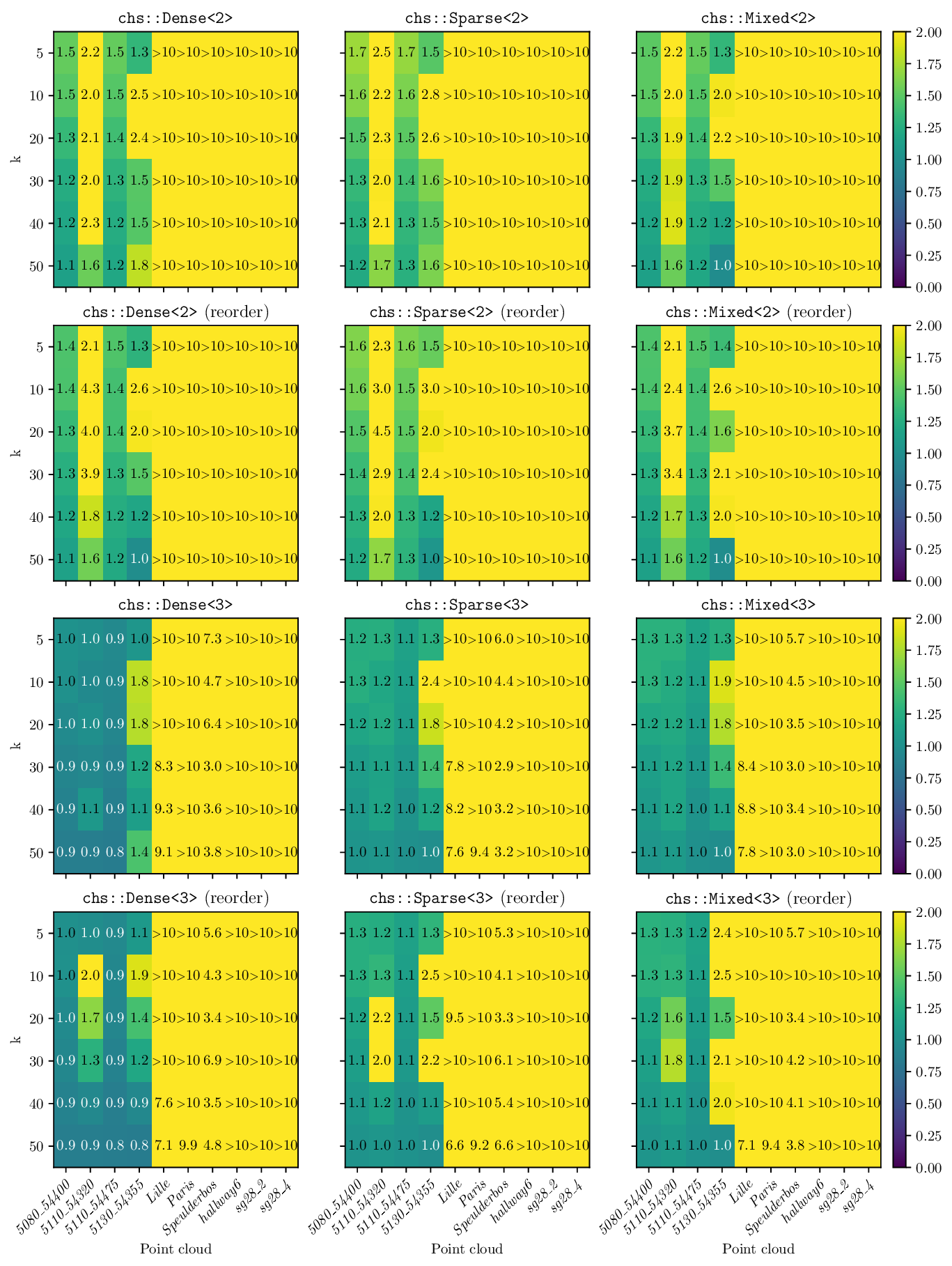}
  \caption{Normalized search time of the \Cheesemap (with cell size \qty{1.0}{\meter}) against \nanoflann in the \KNN search. Lower is better.}
  \label{fig:geom-avg-vs-nanoflann-knn}
\end{figure*}

For the sake of brevity and simplicity, the reported performance metrics correspond to the geometric average of the speedup of the \Cheesemap against \nanoflann across all datasets described in \cref{tab:datasets}.
The geometric average is used as it is the only correct mean when averaging normalized data~\cite{geometric-mean}.
Here, \nanoflann serves as a reference since it is one of the fastest libraries for point cloud indexing and implements all query types used in this study.
\Cref{fig:geom-avg-vs-nanoflann-sphere,fig:geom-avg-vs-nanoflann-cube,fig:geom-avg-vs-nanoflann-knn} show the results for the spherical-kernel search, cubic-kernel search, and \KNN search times, respectively.

The results show that several parameters affect the performance of the \Cheesemap{}: the cell size (\qtylist[list-final-separator={, or }]{1.0;2.5;5.0;7.5;10.0}{\meter}),
the number of dimensions of the grid (2 or 3), the type of kernel (spherical, cubic, or \KNN), and the reordering of the points (enabled or disabled).

\paragraph{Cell size}
The cell size is a key parameter in the performance of the \Cheesemap.
The results show that the performance is better with a smaller cell size, regardless of the type of query.
This is because the cell size determines the granularity of the search, and the smaller the cell size, the more accurate the indexing is and the less time is spent computing distances of points that are not within the search kernel.
Nevertheless, higher performance comes at the expense of a higher memory footprint, as the number of voxels increases, especially for the \chsDense version (see \cref{fig:cheesemap-memory-footprint}).

\paragraph{Dimensionality}
We observed that the performance of the \Cheesemap is highly dependent on the number of dimensions of the grid.
Generally, the performance is better in 3D than in 2D, at the expense of a higher memory footprint.
This is especially noticeable when the cell size is small, as the number of voxels increases.
It is worth noting that the differences are smaller in the \gls{ALS} datasets, where the $z$ dimension is usually smaller than the $x$ and $y$ dimensions (see \cref{fig:search-time-sphere,fig:search-time-cube,fig:search-time-knn}).

\paragraph{Reordering of the points}
The effect of this operation is highly dependent on the point cloud being used.
It was observed that the reordering of the points is beneficial for small point clouds, but its impact is either negligible or counterproductive (contrary to expectations) for larger point clouds.

\paragraph{Kernel-based search}
The type of query does not significantly affect the performance of the \Cheesemap.
The results show that query times are similar for both spherical and cubic kernels, with the cubic kernel performing slightly better.
Search radius affects the performance, resulting in higher query times as the search radius increases since more voxels need to be visited.
It is worth noting that search radii and cell sizes are interrelated, as indexing granularity plays a key role in the \Cheesemap's performance.
For example, performance degradation is observed with small radii and large cell sizes.

\paragraph{\KNN search}
The \KNN search exhibits different behavior from the kernel-based search, with point density being a key factor.
As shown in \cref{fig:geom-avg-vs-nanoflann-knn}, the \Cheesemap's performance is competitive with (and sometimes superior to) \nanoflann for ALS datasets.
However, the performance degrades significantly in other datasets.
Performance could potentially be improved by using smaller cell sizes, at the cost of increasing the memory overhead.

\subsection{Memory footprint of \Cheesemap} \label{sec:cheesemap-memory-footprint}
\begin{figure*}[tb]
  \centering
  \includegraphics[width=\linewidth]{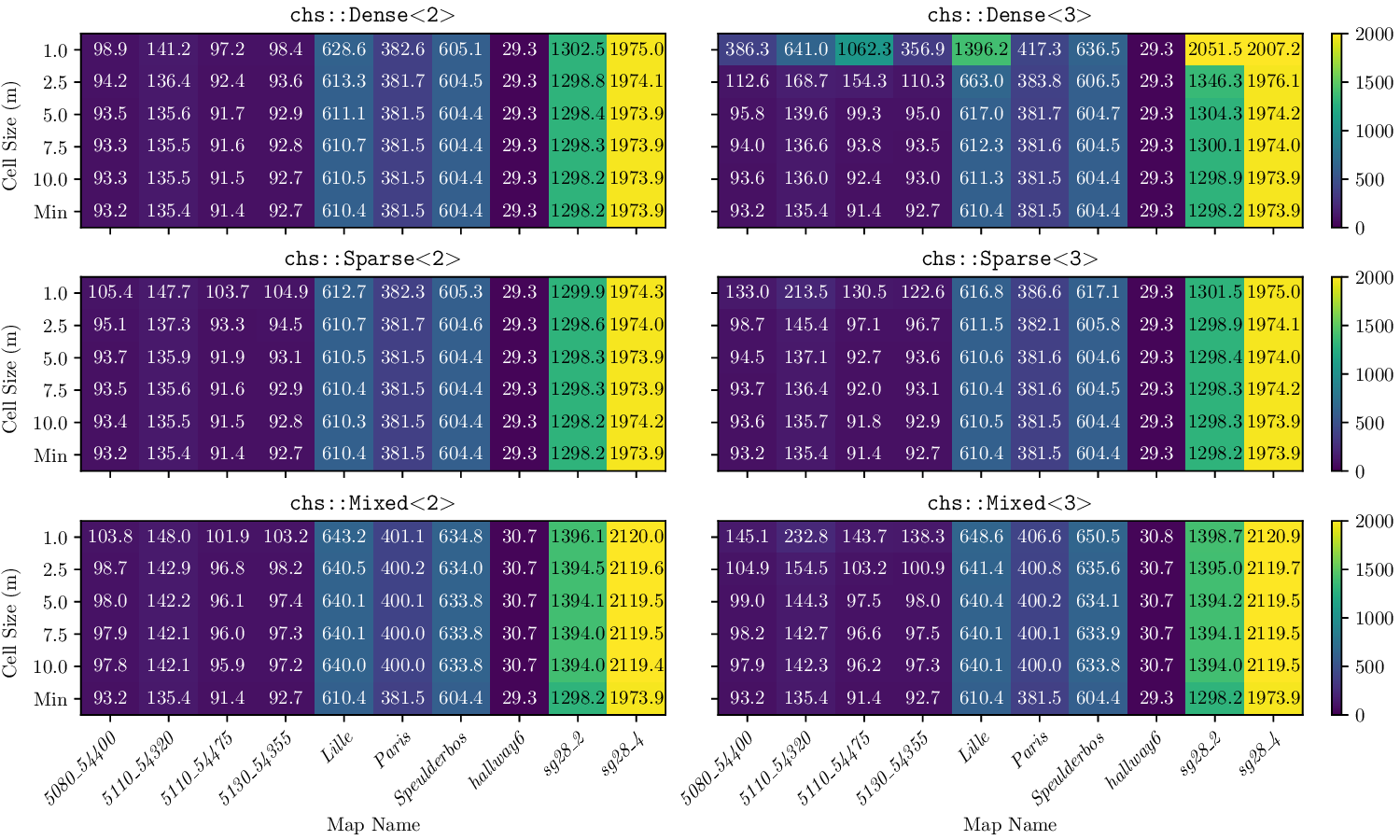}
  \caption{Memory footprint (\unit{\mebi\byte}) of the different \Cheesemap structures for different cell sizes in each dataset and theoretical minimum calculated as the number of points multiplied by the size of a pointer (assumed \qty{8}{\byte}).}
  \label{fig:cheesemap-memory-footprint}
\end{figure*}

Along with the execution time, memory footprint is a key aspect to consider when designing a data structure.
\cref{fig:cheesemap-memory-footprint} shows the memory footprint, as reported by Malt~\cite{Malt2017}, of the different types of \Cheesemap and cell sizes for the datasets described in \cref{tab:datasets}.
This figure also shows the theoretical minimum footprint, calculated assuming an ideal data structure that only needs pointers of $b$ bytes to the $n$ points of the dataset.
In a \num{64}-bit system, the expected minimum overhead would be $8n$ bytes.

Note that the memory footprint is the sum of two elements:
First, the memory needed to store the pointers to each point of the dataset, which depends on the size of the dataset;
Second, the memory needed to store the \Cheesemap itself, which essentially includes the voxels where the point pointers are stored.
Since the different types of \Cheesemap have to store the same number of points, the difference in memory usage is exclusively due to the number of voxels created in each case.

The difference between the two- and three-dimensional versions of the \Cheesemap is clear, with the memory footprint increasing as the cell size decreases.
This difference is more noticeable in the \chsDense case, since all the space is filled with voxels.

Regarding the different types of \Cheesemap, the largest memory footprint is found in the \chsDense version.
This is expected since all the space is filled with voxels, even if they are empty.
The \chsSparse version has a lower memory footprint since only the voxels that contain points are created.
It is worth noting that the \chsMixed is always close to the most memory-efficient version, meeting its design goal.
In some cases, the memory footprint of the \chsMixed is lower than the \chsSparse because the \chsMixed stores some slices of the grid in a dense way.
This is better, memory-wise, than using the sparse representation with a high density of non-empty voxels.

The larger the cell size, the smaller the difference in memory footprint between the two- and three-dimensional versions of the \Cheesemap, since typically the extent of the $z$ dimension is negligible compared with the other two dimensions.
This is especially true for the \gls{ALS} datasets.

There are two datasets where the difference between \chsDense[2] and \chsDense[3] is more evident: \dalesC and \Lille.
These datasets show some peculiarities that are worth mentioning.

\begin{figure}[tb]
  \centering
  \includegraphics[width=\linewidth]{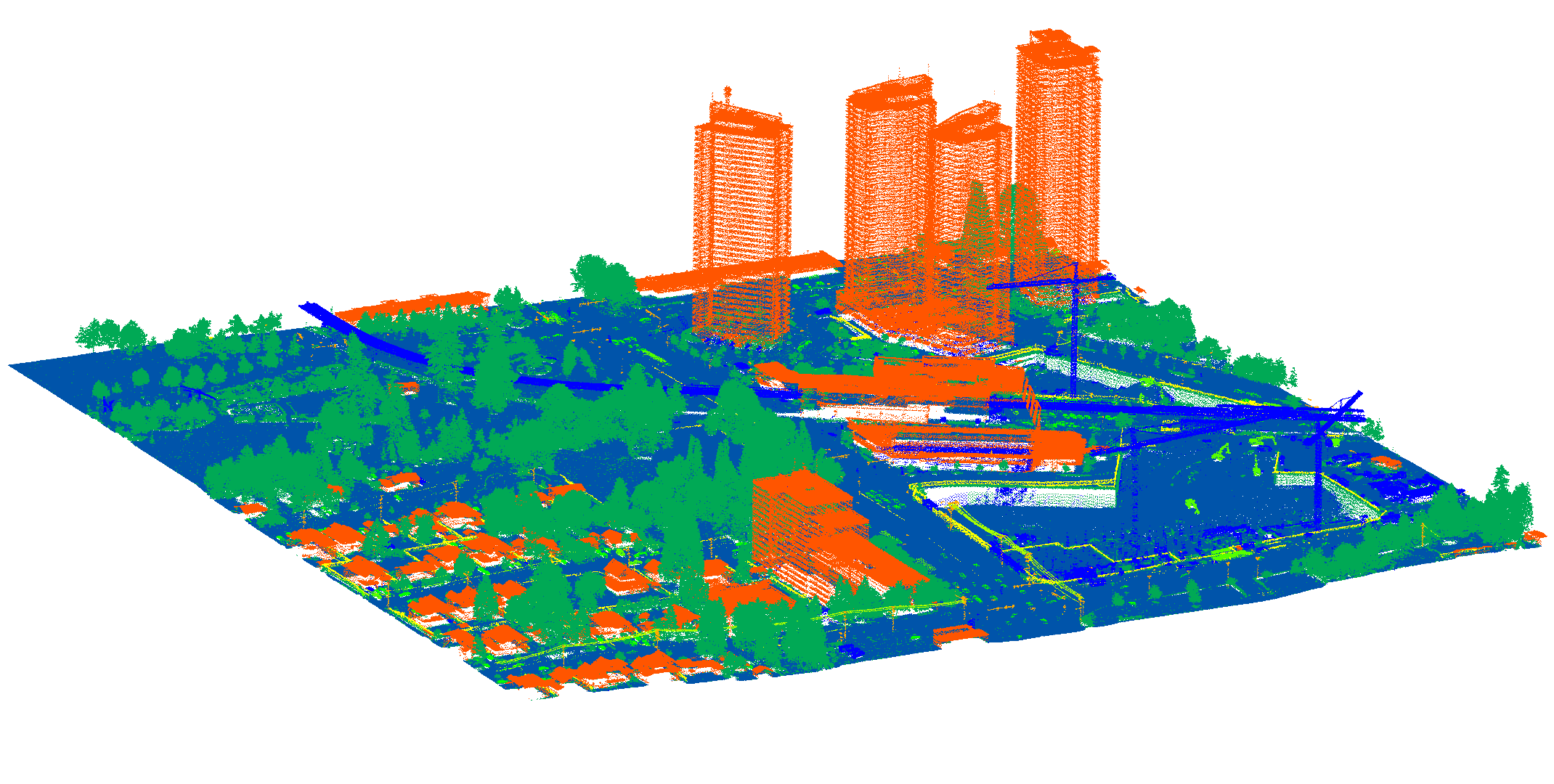}
  \caption{Point cloud of \dalesC dataset.}
  \label{fig:dalesC-pointcloud}
\end{figure}

In the case of \dalesC, the memory footprint of the \chsDense[3] case is an order of magnitude higher than the corresponding two-dimensional case.
As shown in \cref{fig:dalesC-pointcloud}, the presence of skyscrapers in the scene causes the $Z$ dimension of the bounding box to be large compared to the other datasets extracted from DALES.
Because of this, the volume of the bounding box is large, and the number of voxels created increases accordingly.
Quantitatively, the dimensions of the bounding box are \qtyproduct{500 x 500 x 168.4}{\meter}.
When the cell size is \qty{1}{\meter}, the \chsDense[3] creates \num{42.42e6} voxels, of which \qty{98.25}{\percent} are empty. 
In the case of \chsDense[2], the number of created voxels is \num{251.00e3}, of which only \num{428} (\qty{0.17}{\percent}) are empty.

\begin{figure}[tb]
  \centering
  \includegraphics[width=\linewidth]{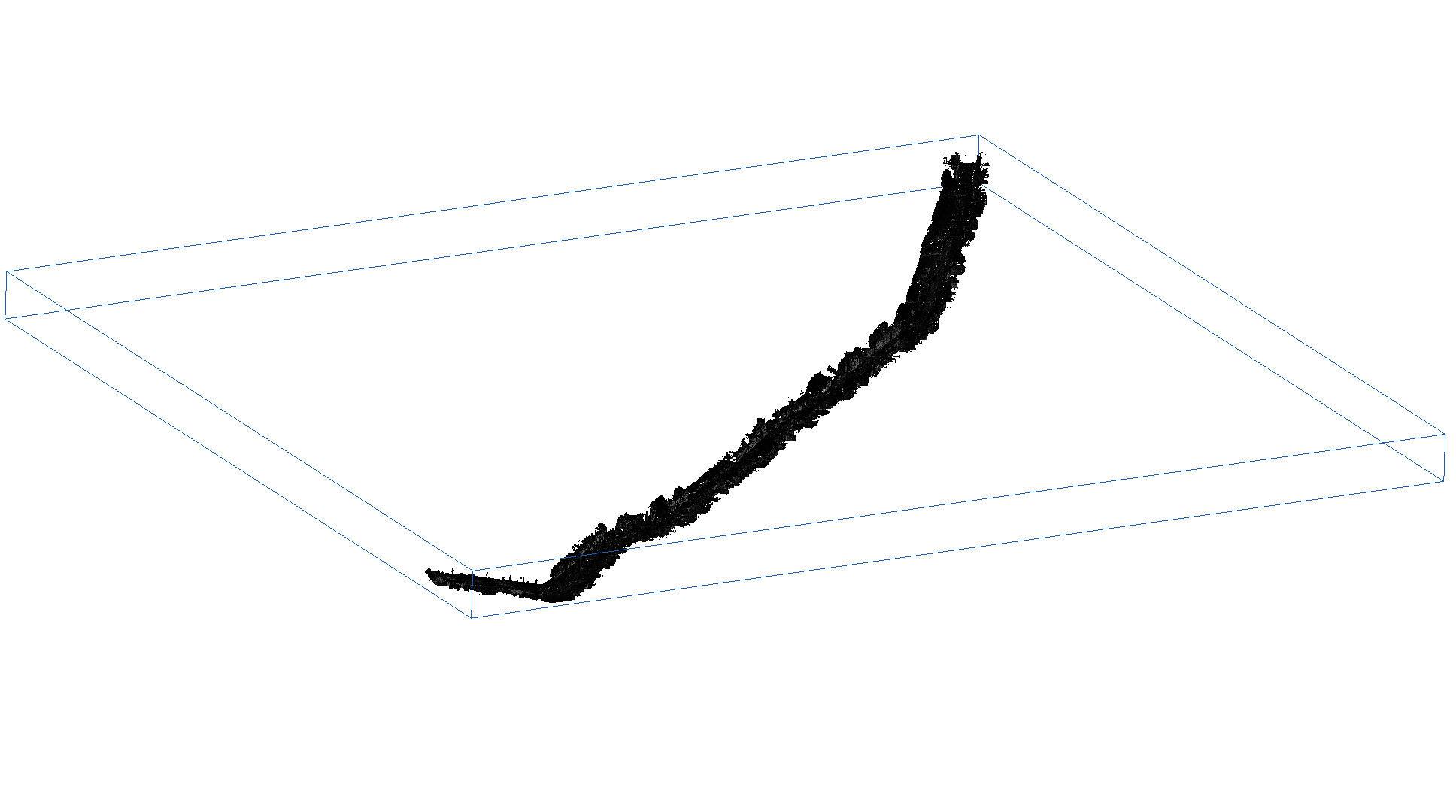}
  \caption{Point cloud of \Lille dataset.}
  \label{fig:lille-pointcloud}
\end{figure}

As for the \Lille dataset, the memory footprint in the \chsDense[3] case is about \num{2.27} times higher compared to the \chsDense[2] case.
A representation of the point cloud with the corresponding bounding box is shown in \cref{fig:lille-pointcloud}.
As we can observe, the \Lille dataset represents a long, quasi-linear street.
The specific alignment of the dataset with the axes results in the largest possible bounding box among all orientations.
Consequently, the vast majority of the voxels are empty for this dataset.
In the \chsDense[2] case, \num{798.43e3} voxels are created, of which \qty{94.31}{\percent} are empty. 
In the \chsDense[3] case, \num{34.33e6} voxels are created, with \num{34.20e6} empty voxels, corresponding to \qty{99.62}{\percent} of the total voxels.


\subsection{Comparison against other data structures} \label{sec:perf-comparison}
In this section, we compare the performance of the \Cheesemap against other data structures.
Regarding the \Cheesemap, we use a cell size of \qty{1.0}{\meter} as it provides the best performance without excessively compromising memory overhead.
The rest of the data structures are used with their default parameters.
Note that different parameters (e.g., the cell size) could be used for different datasets but, for fairness, we have used the same parameters across all datasets.
Similarly, reordering of points is not used in the \Cheesemap to ensure a fair comparison.

\begin{figure*}[ptb]
  \centering
  \includegraphics[width=\linewidth]{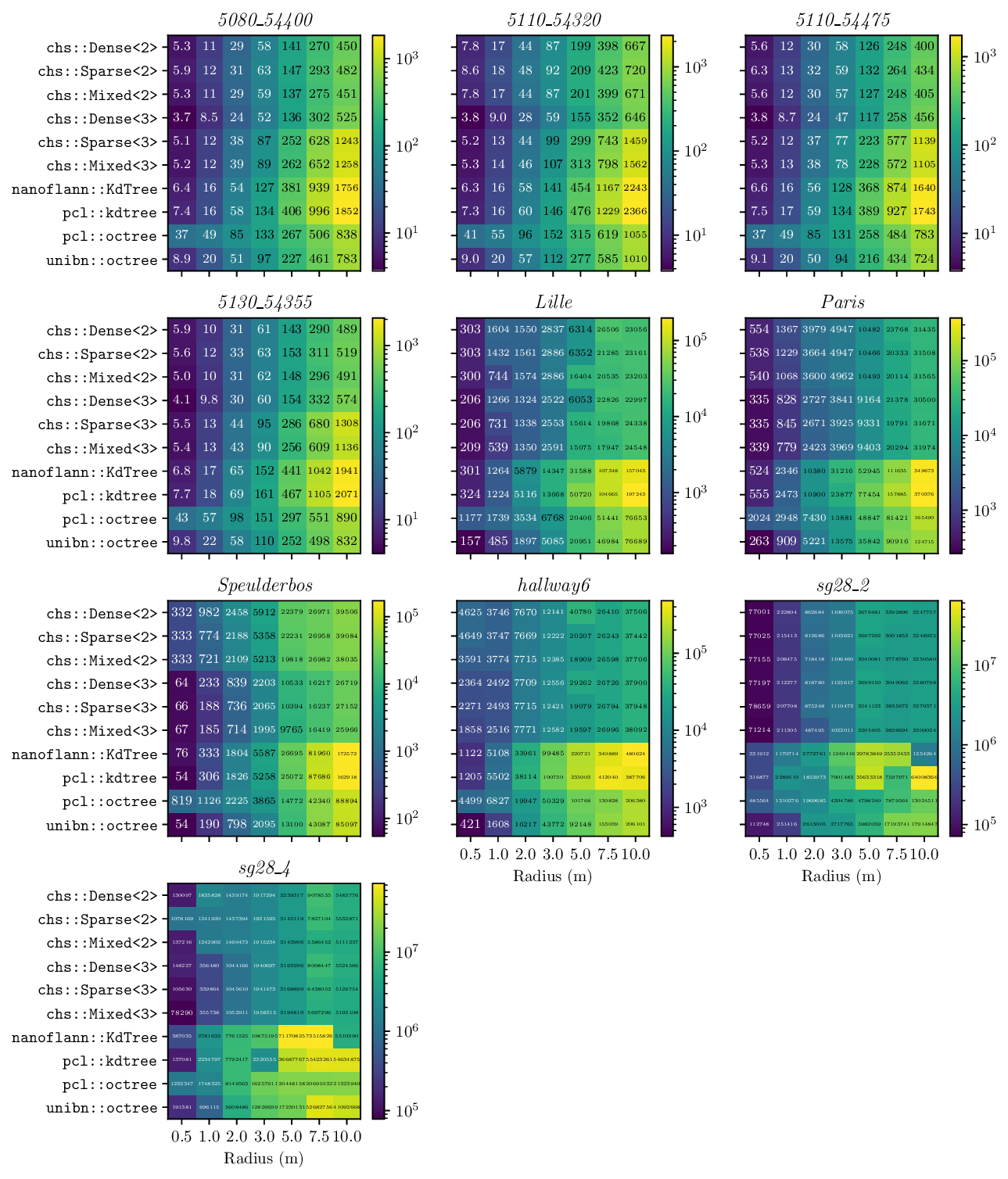}
  \caption{Search time (in \si{\micro\second}) of the different data structures in the spherical-kernel search. Lower is better.}
  \label{fig:search-time-sphere}
\end{figure*}

\begin{figure*}[ptb]
  \centering
  \includegraphics[width=\linewidth]{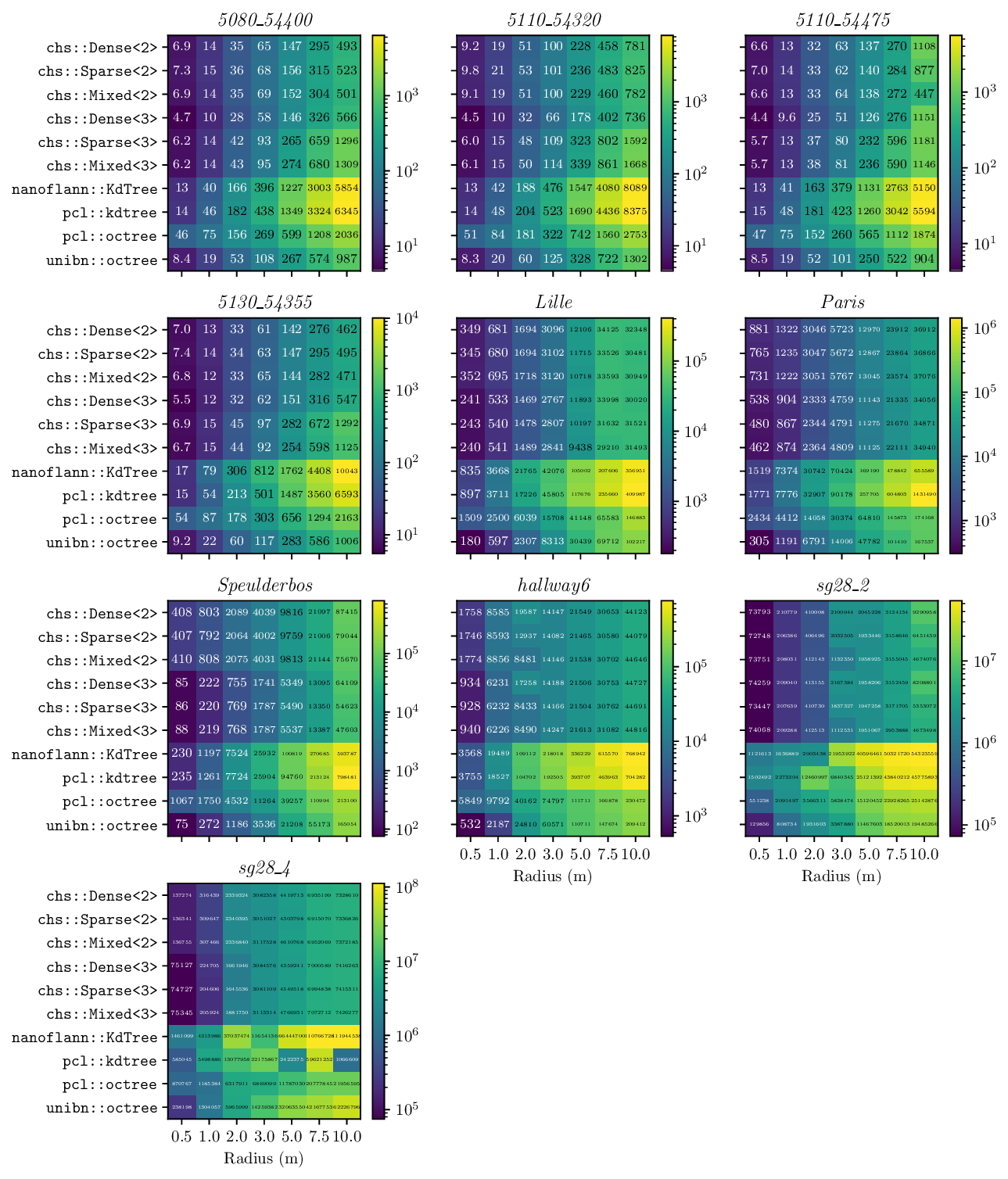}
  \caption{Search time (in \si{\micro\second}) of the different data structures in the cube-kernel search. Lower is better.}
  \label{fig:search-time-cube}
\end{figure*}

\begin{figure*}[ptb]
  \centering
  \includegraphics[width=\linewidth]{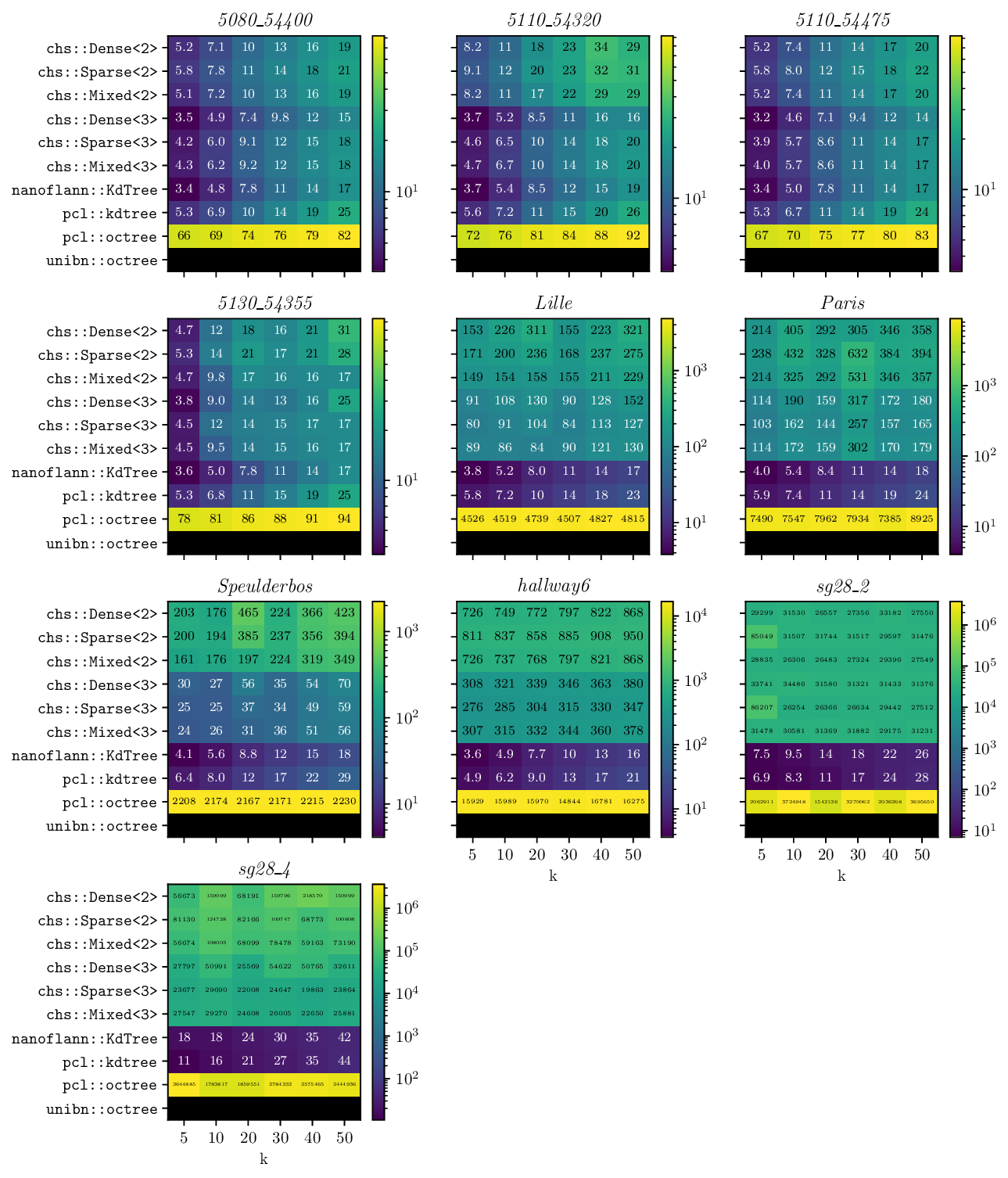}
  \caption{Search time (in \si{\micro\second}) of the different data structures in the \KNN search. \unibnOctree in black since it does not implement this operation. Lower is better.}
  \label{fig:search-time-knn}
\end{figure*}

Results for spherical-, cubic-kernel search, and \KNN search are shown in \cref{fig:search-time-sphere,,fig:search-time-cube,,fig:search-time-knn}, respectively.
The memory footprint of the different data structures is shown in \cref{tab:memory-footprint}.

\paragraph{Kernel search}
In the fixed-radius search, two scenarios can be observed.
For small radii, the performance is highly dependent on the indexing structure and how fast a structure can retrieve points within a small region.
For large radii, the performance depends more on how fast the structure can traverse its internal structure to find points within the radius.
The \pclOctree exemplifies this phenomenon since for small radii it is the slowest data structure, but as the radius increases it becomes competitive with other structures.
As a general rule, most structures have comparable performance with small radii, but performance differences favor the \Cheesemap variants as the radius increases.

Additionally, some observations can be made regarding different dataset types.
For ALS point clouds with spherical- and cubic-kernel search, the \chsDense[3] is the fastest data structure, followed by \chsSparse[3] and \chsMixed[3].
The \chsDense[3] achieves up to \qty{45}{\percent} better performance than \nanoflann.

For MLS point clouds (\Lille and \Paris), the \unibnOctree is the fastest data structure, followed by the three-dimensional versions of \Cheesemap.
The high point cloud density and relatively large cell size (even at \qty{1.0}{\cubic\meter}) make the \Cheesemap slower than the \unibnOctree.
We hypothesize that a smaller cell size could improve the performance of the \Cheesemap in these datasets.

In the \Speulderbos dataset (ULS), there is a significant difference between two- and three-dimensional \Cheesemap versions, with the latter performing almost on par with \unibnOctree and \nanoflann.
The $z$ dimension is significant due to tree presence, causing many points to stack along the vertical axis.
Therefore, considering the $z$ axis benefits the \Cheesemap for this dataset.

Unlike other datasets, for the \Hallway dataset (TLS), the \Cheesemap is slower for very small radii queries but becomes the fastest as radius increases.
This results from the small dimensions (and high density) of the dataset, where a \qty{1.0}{\meter} cell size is too large for good performance with small radii queries.
When the radius exceeds the cell size, the three-dimensional \Cheesemap versions become the fastest structures.

Finally, for the \sgRural and \sgUrban datasets (TLS), the \Cheesemap is the fastest data structure, with the \chsMixed[3] being the fastest version.
These results outline the benefits of the \Cheesemap, considering that these are the largest datasets in terms of the number of points.

\paragraph{\KNN search}
Before commenting on the results, it should be mentioned that the \unibnOctree does not support \KNN search, so it is not included in the comparison.

As mentioned in \cref{sec:cheesemap-performance}, the \Cheesemap shows notable differences between datasets for \KNN search.
For ALS datasets, the \Cheesemap performs on par with \nanoflann and \pclkdtree, with \nanoflann and \chsDense[3] being slightly faster than the others.
For the rest of the datasets, \nanoflann is the fastest data structure, closely followed by \pclkdtree.
The \Cheesemap is significantly slower, but still faster than the \pclOctree, which is the slowest data structure across all point clouds.

These results highlight the importance of point density in the performance of the \Cheesemap.
While a dataset-specific cell size might improve its performance for \KNN search, this would require more detailed analysis and would reduce the generalizability of the results.

\paragraph{Memory footprint}
As mentioned in \cref{sec:cheesemap-memory-footprint}, the memory footprint of indexing data structures is composed of two elements: the memory needed to store the pointers to the points and the memory needed to store the data structure itself.
\Cref{tab:memory-footprint} shows the memory footprint of the different data structures as reported by Malt~\cite{Malt2017} alongside the theoretical minimum overhead.

The results show that two structures have the lowest memory footprint: \nanoflann and \chsDense[2].
All two-dimensional versions of the \Cheesemap have a memory footprint close to the theoretical minimum.
The three-dimensional versions of the \Cheesemap have a slightly higher memory footprint, around \qty{30}{\percent} higher than the minimum.
As mentioned in \cref{sec:cheesemap-memory-footprint}, the \chsDense[3] has a higher memory footprint but remains comparable to other state-of-the-art data structures.

Finally, the \pclOctree and \unibnOctree are generally close to each other but use approximately twice the memory of \nanoflann.
The \pclkdtree has the highest memory footprint, using up to \num{6} times more memory than \nanoflann or \Cheesemap in the \sgRural dataset.

\begin{table*}[tb]
  \centering
  \caption{Memory footprint (\unit{\mebi\byte}) for the different data structures in each dataset and theoretical minimum calculated as the number of points multiplied by the size of a pointer (assumed \qty{8}{\byte}). Lower is better.}
  \label{tab:memory-footprint}
  \begin{adjustbox}{max width=\linewidth}
    \begin{tabular}{l*{10}{S[table-format=4.1]}}
      \toprule
      Map            & {\dalesA} & {\dalesB} & {\dalesC} & {\dalesD} & {\Lille} & {\Paris} & {\Speulderbos} & {\Hallway} & {\sgRural} & {\sgUrban} \\
      Points (M)     & 12.2      & 17.7      & 11.9      & 12.1      & 80.0     & 50.0     & 79.2           & 3.8        & 170.2      & 258.7      \\
      Theor. minimum & 93.2      & 135.4     & 91.4      & 92.7      & 610.3    & 381.4    & 604.4          & 29.3       & 1298.2     & 1973.9     \\
      \midrule
      \chsDense[2]   & 98.9      & 141.2     & 97.2      & 98.4      & 628.6    & 382.6    & 605.1          & 29.3       & 1302.5     & 1975.0     \\
      \chsDense[3]   & 386.3     & 641.0     & 1073.5    & 356.9     & 1396.2   & 417.3    & 636.5          & 29.3       & 2051.5     & 2007.2     \\
      \chsSparse[2]  & 105.4     & 147.7     & 103.7     & 104.9     & 612.7    & 382.3    & 605.3          & 29.3       & 1299.9     & 1974.3     \\
      \chsSparse[3]  & 133.0     & 213.5     & 130.5     & 122.6     & 616.8    & 386.6    & 617.1          & 29.3       & 1301.5     & 1975.0     \\
      \chsMixed[2]   & 103.8     & 148.0     & 101.9     & 103.2     & 643.2    & 401.1    & 634.8          & 30.7       & 1396.1     & 2120.0     \\
      \chsMixed[3]   & 145.1     & 232.8     & 413.7     & 138.3     & 648.6    & 406.6    & 650.5          & 30.8       & 1398.7     & 2120.9     \\
      \midrule
      \nanoflann     & 98.9      & 141.7     & 94.9      & 92.8      & 672.8    & 381.5    & 676.3          & 29.3       & 1327.3     & 1970.8     \\
      \pclkdtree     & 580.4     & 840.9     & 569.4     & 579.1     & 3779.3   & 2373.2   & 3769.3         & 180.4      & 8032.2     & 12208.8    \\
      \pclOctree     & 186.4     & 270.8     & 182.8     & 185.3     & 1259.7   & 783.5    & 1171.8         & 58.6       & 2596.4     & 2817.3     \\
      \unibnOctree   & 174.4     & 254.6     & 176.4     & 182.9     & 1122.6   & 717.0    & 1056.9         & 50.0       & 2226.1     & 3381.6     \\
      \bottomrule
    \end{tabular}
  \end{adjustbox}
\end{table*}

\section{Conclusions}\label{sec:conclusions}
In this paper, we presented a thorough analysis of several data structures for indexing point clouds and compared them in terms of performance and memory footprint in real-world datasets.
We also introduced the \Cheesemap, a novel data structure designed to be fast and memory-efficient for LiDAR point clouds.
The \Cheesemap is based on a grid structure that divides the space into voxels, where each voxel lists the points within it.
This structure allows for fast queries of points within a certain radius (spherical- and cubic-kernel search) and \KNN search.

The novelty of the \Cheesemap lies in its three indexing strategies: dense, sparse, and mixed.
The dense representation creates a voxel for each cell, regardless of whether it contains points.
The sparse structure only creates voxels for cells containing points, storing them in a hashmap.
The mixed strategy is a hybrid approach where the grid is divided into slices, each being dense or sparse depending on its non-empty voxel distribution.

The results show that the \Cheesemap's performance depends heavily on cell size and dimensionality, being best with small cell sizes and three-dimensional grids at the cost of a slightly higher memory footprint.

Compared to other state-of-the-art data structures, the \Cheesemap shows superior performance for \gls{ALS} and large point clouds for fixed-radius searches, being up to \qty{45}{\percent} faster than \nanoflann.
For \KNN, the performance is similar to \nanoflann, the fastest library that implements this type of query. Regarding datasets acquired using other kinds of platforms besides \gls{ALS}, the performance for fixed-radius searches is on par with that of the fastest libraries, while the go-to option for \KNN searches, in this case, is \nanoflann.
Finally, the memory footprint of the \Cheesemap is close to the theoretical minimum overhead, which is shown to be more efficient than \pclkdtree, \pclOctree, and \unibnOctree.
These facts make the \Cheesemap the best data structure for performing different types of queries across several types of datasets.

In future work, we plan to improve the \Cheesemap by exploring alternatives to \verb|std::unordered_map| to enhance the performance of sparse and mixed representations.
Additional improvements could come from optimizing \KNN search and developing automatic mechanisms for selecting optimal cell sizes for any given point cloud.

\bibliographystyle{elsarticle-num}
\bibliography{biblio}

\begin{thebibliography}{10}
\expandafter\ifx\csname url\endcsname\relax
  \def\url#1{\texttt{#1}}\fi
\expandafter\ifx\csname urlprefix\endcsname\relax\def\urlprefix{URL }\fi
\expandafter\ifx\csname href\endcsname\relax
  \def\href#1#2{#2} \def\path#1{#1}\fi

\bibitem{Xiang2022}
Q.~Xiang, Y.~He, D.~Wen, Adaptive deep learning-based neighborhood search
  method for point cloud, Scientific Reports (2022).
\newblock \href {https://doi.org/10.1038/s41598-022-06200-z}
  {\path{doi:10.1038/s41598-022-06200-z}}.

\bibitem{Turau1991}
V.~Turau, Fixed-radius near neighbors search, Information Processing Letters
  (1991).
\newblock \href {https://doi.org/10.1016/0020-0190(91)90180-p}
  {\path{doi:10.1016/0020-0190(91)90180-p}}.

\bibitem{Fix1989}
E.~Fix, J.~L. Hodges,
  \href{http://www.jstor.org/stable/1403797}{{Discriminatory Analysis.
  Nonparametric Discrimination: Consistency Properties}}, International
  Statistical Review / Revue Internationale de Statistique 57~(3) (1989)
  238--247.
\newblock \href {https://doi.org/10.2307/1403797} {\path{doi:10.2307/1403797}}.
\newline\urlprefix\url{http://www.jstor.org/stable/1403797}

\bibitem{Cover1967}
T.~Cover, P.~Hart, Nearest neighbor pattern classification, IEEE Transactions
  on Information Theory 13~(1) (1967) 21--27.
\newblock \href {https://doi.org/10.1109/TIT.1967.1053964}
  {\path{doi:10.1109/TIT.1967.1053964}}.

\bibitem{Filin2005}
S.~Filin, N.~Pfeifer,
  \href{https://www.scopus.com/inward/record.uri?eid=2-s2.0-27444443641&doi=10.14358%2fPERS.71.6.743&partnerID=40&md5=20f5ee6ed3d6c95e80b3308e85e00007}{Neighborhood
  systems for airborne laser data}, Photogrammetric Engineering and Remote
  Sensing 71~(6) (2005) 743 – 755.
\newblock \href {https://doi.org/10.14358/PERS.71.6.743}
  {\path{doi:10.14358/PERS.71.6.743}}.
\newline\urlprefix\url{https://www.scopus.com/inward/record.uri?eid=2-s2.0-27444443641&doi=10.14358%2fPERS.71.6.743&partnerID=40&md5=20f5ee6ed3d6c95e80b3308e85e00007}

\bibitem{Wang2019}
L.~Wang, Y.~Huang, Y.~Hou, S.~Zhang, J.~Shan, Graph attention convolution for
  point cloud semantic segmentation, in: 2019 IEEE/CVF Conference on Computer
  Vision and Pattern Recognition (CVPR), 2019, pp. 10288--10297.
\newblock \href {https://doi.org/10.1109/CVPR.2019.01054}
  {\path{doi:10.1109/CVPR.2019.01054}}.

\bibitem{Thomas2018}
H.~Thomas, F.~Goulette, J.-E. Deschaud, B.~Marcotegui, Y.~LeGall, Semantic
  classification of {3D} point clouds with multiscale spherical neighborhoods,
  in: 2018 International Conference on 3D Vision (3DV), 2018, pp. 390--398.
\newblock \href {https://doi.org/10.1109/3DV.2018.00052}
  {\path{doi:10.1109/3DV.2018.00052}}.

\bibitem{Weinmann2015}
M.~Weinmann, B.~Jutzi, S.~Hinz, C.~Mallet,
  \href{https://www.sciencedirect.com/science/article/pii/S0924271615000349}{Semantic
  point cloud interpretation based on optimal neighborhoods, relevant features
  and efficient classifiers}, ISPRS Journal of Photogrammetry and Remote
  Sensing 105 (2015) 286--304.
\newblock \href {https://doi.org/10.1016/j.isprsjprs.2015.01.016}
  {\path{doi:10.1016/j.isprsjprs.2015.01.016}}.
\newline\urlprefix\url{https://www.sciencedirect.com/science/article/pii/S0924271615000349}

\bibitem{Hackel2016}
T.~Hackel, J.~D. Wegner, K.~Schindler,
  \href{https://isprs-annals.copernicus.org/articles/III-3/177/2016/}{Fast
  semantic segmentation of {3D} point clouds with strongly varying density},
  ISPRS Annals of the Photogrammetry, Remote Sensing and Spatial Information
  Sciences III-3 (2016) 177--184.
\newblock \href {https://doi.org/10.5194/isprs-annals-III-3-177-2016}
  {\path{doi:10.5194/isprs-annals-III-3-177-2016}}.
\newline\urlprefix\url{https://isprs-annals.copernicus.org/articles/III-3/177/2016/}

\bibitem{Pauly2003}
M.~Pauly, R.~Keiser, M.~Gross,
  \href{https://onlinelibrary.wiley.com/doi/abs/10.1111/1467-8659.00675}{Multi-scale
  feature extraction on point-sampled surfaces}, Computer Graphics Forum 22~(3)
  (2003) 281--289.
\newblock \href
  {http://arxiv.org/abs/https://onlinelibrary.wiley.com/doi/pdf/10.1111/1467-8659.00675}
  {\path{arXiv:https://onlinelibrary.wiley.com/doi/pdf/10.1111/1467-8659.00675}},
  \href {https://doi.org/10.1111/1467-8659.00675}
  {\path{doi:10.1111/1467-8659.00675}}.
\newline\urlprefix\url{https://onlinelibrary.wiley.com/doi/abs/10.1111/1467-8659.00675}

\bibitem{Brodu2012}
N.~Brodu, D.~Lague,
  \href{https://www.sciencedirect.com/science/article/pii/S0924271612000330}{{3D}
  terrestrial {LiDAR} data classification of complex natural scenes using a
  multi-scale dimensionality criterion: Applications in geomorphology}, ISPRS
  Journal of Photogrammetry and Remote Sensing 68 (2012) 121--134.
\newblock \href {https://doi.org/10.1016/j.isprsjprs.2012.01.006}
  {\path{doi:10.1016/j.isprsjprs.2012.01.006}}.
\newline\urlprefix\url{https://www.sciencedirect.com/science/article/pii/S0924271612000330}

\bibitem{Niemeyer2014}
J.~Niemeyer, F.~Rottensteiner, U.~Soergel,
  \href{https://www.sciencedirect.com/science/article/pii/S0924271613002359}{Contextual
  classification of lidar data and building object detection in urban areas},
  ISPRS Journal of Photogrammetry and Remote Sensing 87 (2014) 152--165.
\newblock \href {https://doi.org/10.1016/j.isprsjprs.2013.11.001}
  {\path{doi:10.1016/j.isprsjprs.2013.11.001}}.
\newline\urlprefix\url{https://www.sciencedirect.com/science/article/pii/S0924271613002359}

\bibitem{Cook1986}
R.~L. Cook, \href{https://doi.org/10.1145/7529.8927}{Stochastic sampling in
  computer graphics}, ACM Trans. Graph. 5~(1) (1986) 51–72.
\newblock \href {https://doi.org/10.1145/7529.8927}
  {\path{doi:10.1145/7529.8927}}.
\newline\urlprefix\url{https://doi.org/10.1145/7529.8927}

\bibitem{Sevgen2023}
E.~Sevgen, S.~Abdikan,
  \href{https://www.mdpi.com/2072-4292/15/15/3787}{Classification of
  large-scale mobile laser scanning data in urban area with lightgbm}, Remote
  Sensing 15~(15) (2023).
\newblock \href {https://doi.org/10.3390/rs15153787}
  {\path{doi:10.3390/rs15153787}}.
\newline\urlprefix\url{https://www.mdpi.com/2072-4292/15/15/3787}

\bibitem{Tian2020}
Y.~Tian, W.~Song, L.~Chen, Y.~Sung, J.~Kwak, S.~Sun,
  \href{https://www.mdpi.com/1424-8220/20/8/2309}{A fast spatial clustering
  method for sparse lidar point clouds using gpu programming}, Sensors 20~(8)
  (2020).
\newblock \href {https://doi.org/10.3390/s20082309}
  {\path{doi:10.3390/s20082309}}.
\newline\urlprefix\url{https://www.mdpi.com/1424-8220/20/8/2309}

\bibitem{Zeng2009}
X.~Zeng, W.~He, \href{https://doi.org/10.1117/12.833740}{{GPGPU-based parallel
  processing of massive LiDAR point cloud}}, in: F.~Zhang, F.~Zhang (Eds.),
  MIPPR 2009: Medical Imaging, Parallel Processing of Images, and Optimization
  Techniques, Vol. 7497, International Society for Optics and Photonics, SPIE,
  2009, p. 749716.
\newblock \href {https://doi.org/10.1117/12.833740}
  {\path{doi:10.1117/12.833740}}.
\newline\urlprefix\url{https://doi.org/10.1117/12.833740}

\bibitem{Li2021}
Q.~Li, P.~Yuan, Y.~Lin, Y.~Tong, X.~Liu,
  \href{https://doi.org/10.1117/1.JRS.15.024523}{{Pointwise classification of
  mobile laser scanning point clouds of urban scenes using raw data}}, Journal
  of Applied Remote Sensing 15~(2) (2021) 024523.
\newblock \href {https://doi.org/10.1117/1.JRS.15.024523}
  {\path{doi:10.1117/1.JRS.15.024523}}.
\newline\urlprefix\url{https://doi.org/10.1117/1.JRS.15.024523}

\bibitem{Bentley1975}
J.~L. Bentley, \href{https://doi.org/10.1145/361002.361007}{Multidimensional
  binary search trees used for associative searching}, Commun. ACM 18~(9)
  (1975) 509--517.
\newblock \href {https://doi.org/10.1145/361002.361007}
  {\path{doi:10.1145/361002.361007}}.
\newline\urlprefix\url{https://doi.org/10.1145/361002.361007}

\bibitem{Behley2015}
J.~Behley, V.~Steinhage, A.~B. Cremers, Efficient radius neighbor search in
  three-dimensional point clouds, in: 2015 IEEE International Conference on
  Robotics and Automation (ICRA), 2015, pp. 3625--3630.
\newblock \href {https://doi.org/10.1109/ICRA.2015.7139702}
  {\path{doi:10.1109/ICRA.2015.7139702}}.

\bibitem{Omohundro1989}
S.~M. Omohundro,
  \href{https://steveomohundro.com/wp-content/uploads/2009/03/omohundro89_five_balltree_construction_algorithms.pdf}{Five
  balltree construction algorithms}, International Computer Science Institute
  Berkeley, 1989.
\newline\urlprefix\url{https://steveomohundro.com/wp-content/uploads/2009/03/omohundro89_five_balltree_construction_algorithms.pdf}

\bibitem{Guttman1984}
A.~Guttman, \href{https://doi.org/10.1145/602259.602266}{R-trees: a dynamic
  index structure for spatial searching}, in: Proceedings of the 1984 ACM
  SIGMOD International Conference on Management of Data, SIGMOD '84,
  Association for Computing Machinery, New York, NY, USA, 1984, p. 47–57.
\newblock \href {https://doi.org/10.1145/602259.602266}
  {\path{doi:10.1145/602259.602266}}.
\newline\urlprefix\url{https://doi.org/10.1145/602259.602266}

\bibitem{Luo2022}
C.~Luo, X.~Li, N.~Cheng, H.~Li, S.~Lei, P.~Li,
  \href{https://arxiv.org/abs/2201.12769}{{MVP-Net}: Multiple view pointwise
  semantic segmentation of large-scale point clouds} (2022).
\newblock \href {http://arxiv.org/abs/2201.12769} {\path{arXiv:2201.12769}}.
\newline\urlprefix\url{https://arxiv.org/abs/2201.12769}

\bibitem{Che2019}
E.~Che, M.~J. Olsen, \href{https://www.mdpi.com/2072-4292/11/7/836}{An
  efficient framework for mobile lidar trajectory reconstruction and mo-norvana
  segmentation}, Remote Sensing 11~(7) (2019).
\newblock \href {https://doi.org/10.3390/rs11070836}
  {\path{doi:10.3390/rs11070836}}.
\newline\urlprefix\url{https://www.mdpi.com/2072-4292/11/7/836}

\bibitem{Wang2021}
W.~Wang, Y.~Zhang, G.~Ge, Q.~Jiang, Y.~Wang, L.~Hu,
  \href{https://www.mdpi.com/2076-3417/11/20/9581}{A hybrid spatial indexing
  structure of massive point cloud based on octree and {3D R*-Tree}}, Applied
  Sciences 11~(20) (2021).
\newblock \href {https://doi.org/10.3390/app11209581}
  {\path{doi:10.3390/app11209581}}.
\newline\urlprefix\url{https://www.mdpi.com/2076-3417/11/20/9581}

\bibitem{Feng2013}
Y.~Feng, M.~Cen, T.~Zhang, \href{https://doi.org/10.1117/12.2019658}{{An
  algorithm of fast index constructing and neighbor searching for 3D LiDAR
  data}}, in: H.~Tan (Ed.), PIAGENG 2013: Intelligent Information, Control, and
  Communication Technology for Agricultural Engineering, Vol. 8762,
  International Society for Optics and Photonics, SPIE, 2013, p. 87620Y.
\newblock \href {https://doi.org/10.1117/12.2019658}
  {\path{doi:10.1117/12.2019658}}.
\newline\urlprefix\url{https://doi.org/10.1117/12.2019658}

\bibitem{Li2018}
J.~Li, B.~M. Chen, G.~H. Lee, {SO-Net}: Self-organizing network for point cloud
  analysis, in: 2018 IEEE/CVF Conference on Computer Vision and Pattern
  Recognition, 2018, pp. 9397--9406.
\newblock \href {https://doi.org/10.1109/CVPR.2018.00979}
  {\path{doi:10.1109/CVPR.2018.00979}}.

\bibitem{Zhan2018}
X.~Zhan, Y.~Cai, P.~He, \href{https://doi.org/10.1177/1687814018814330}{A
  three-dimensional point cloud registration based on entropy and particle
  swarm optimization}, Advances in Mechanical Engineering 10~(12) (2018)
  1687814018814330.
\newblock \href {http://arxiv.org/abs/https://doi.org/10.1177/1687814018814330}
  {\path{arXiv:https://doi.org/10.1177/1687814018814330}}, \href
  {https://doi.org/10.1177/1687814018814330}
  {\path{doi:10.1177/1687814018814330}}.
\newline\urlprefix\url{https://doi.org/10.1177/1687814018814330}

\bibitem{Liu2021}
D.~Liu, D.~Li, M.~Wang, Z.~Wang,
  \href{https://www.mdpi.com/2220-9964/10/3/127}{{3D} change detection using
  adaptive thresholds based on local point cloud density}, ISPRS International
  Journal of Geo-Information 10~(3) (2021).
\newblock \href {https://doi.org/10.3390/ijgi10030127}
  {\path{doi:10.3390/ijgi10030127}}.
\newline\urlprefix\url{https://www.mdpi.com/2220-9964/10/3/127}

\bibitem{Elseberg2012}
J.~Elseberg, S.~Magnenat, R.~Siegwart, A.~N{\"u}chter,
  \href{https://robotik.informatik.uni-wuerzburg.de/telematics/download/joser2012.pdf}{Comparison
  of nearest-neighbor-search strategies and implementations for efficient shape
  registration}, Journal of Software Engineering for Robotics 3~(1) (2012)
  2--12.
\newline\urlprefix\url{https://robotik.informatik.uni-wuerzburg.de/telematics/download/joser2012.pdf}

\bibitem{Lawson2022}
M.~Lawson, W.~Gropp, J.~Lofstead, Exploring spatial indexing for accelerated
  feature retrieval in {HPC}, in: 2022 22nd IEEE International Symposium on
  Cluster, Cloud and Internet Computing (CCGrid), 2022, pp. 605--614.
\newblock \href {https://doi.org/10.1109/CCGrid54584.2022.00070}
  {\path{doi:10.1109/CCGrid54584.2022.00070}}.

\bibitem{Varney2020}
N.~Varney, V.~K. Asari, Q.~Graehling, {DALES}: A large-scale aerial {LiDAR}
  data set for semantic segmentation, in: Proceedings of the IEEE/CVF
  Conference on Computer Vision and Pattern Recognition Workshops, 2020, pp.
  186--187.
\newblock \href {https://doi.org/10.1109/CVPRW50498.2020.00101}
  {\path{doi:10.1109/CVPRW50498.2020.00101}}.

\bibitem{Roynard2018}
X.~Roynard, J.-E. Deschaud, F.~Goulette, {Paris-Lille-3D}: A large and
  high-quality ground-truth urban point cloud dataset for automatic
  segmentation and classification, The International Journal of Robotics
  Research 37~(6) (2018) 545--557.
\newblock \href {https://doi.org/10.1177/0278364918767506}
  {\path{doi:10.1177/0278364918767506}}.

\bibitem{Brede2019}
B.~Brede, K.~Calders, A.~Lau, P.~Raumonen, H.~M. Bartholomeus, M.~Herold,
  L.~Kooistra,
  \href{https://www.sciencedirect.com/science/article/pii/S0034425719303748}{Non-destructive
  tree volume estimation through quantitative structure modelling: Comparing
  {UAV} laser scanning with terrestrial {LiDAR}}, Remote Sensing of Environment
  233 (2019) 111355.
\newblock \href {https://doi.org/10.1016/j.rse.2019.111355}
  {\path{doi:10.1016/j.rse.2019.111355}}.
\newline\urlprefix\url{https://www.sciencedirect.com/science/article/pii/S0034425719303748}

\bibitem{Armeni2016}
I.~Armeni, O.~Sener, A.~R. Zamir, H.~Jiang, I.~Brilakis, M.~Fischer,
  S.~Savarese, {3D} semantic parsing of large-scale indoor spaces, in: 2016
  IEEE Conference on Computer Vision and Pattern Recognition (CVPR), 2016, pp.
  1534--1543.
\newblock \href {https://doi.org/10.1109/CVPR.2016.170}
  {\path{doi:10.1109/CVPR.2016.170}}.

\bibitem{Hackel2017}
T.~Hackel, N.~Savinov, L.~Ladicky, J.~D. Wegner, K.~Schindler, M.~Pollefeys,
  \href{https://isprs-annals.copernicus.org/articles/IV-1-W1/91/2017/}{Semantic3d.net:
  A new large-scale point cloud classification benchmark} (2017).
\newblock \href {https://doi.org/10.5194/isprs-annals-IV-1-W1-91-2017}
  {\path{doi:10.5194/isprs-annals-IV-1-W1-91-2017}}.
\newline\urlprefix\url{https://isprs-annals.copernicus.org/articles/IV-1-W1/91/2017/}

\bibitem{blanco2014nanoflann}
J.~L. Blanco, P.~K. Rai, nanoflann: a {C}++ header-only fork of {FLANN}, a
  library for nearest neighbor ({NN}) with kd-trees,
  \url{https://github.com/jlblancoc/nanoflann} (2014).

\bibitem{DBLP:conf/visapp/MujaL09}
M.~Muja, D.~G. Lowe,
  \href{https://www.cs.ubc.ca/research/flann/uploads/FLANN/flann_visapp09.pdf}{Fast
  approximate nearest neighbors with automatic algorithm configuration}, in:
  A.~Ranchordas, H.~Ara{\'{u}}jo (Eds.), {VISAPP} 2009 - Proceedings of the
  Fourth International Conference on Computer Vision Theory and Applications,
  Lisboa, Portugal, February 5-8, 2009 - Volume 1, {INSTICC} Press, 2009, pp.
  331--340.
\newline\urlprefix\url{https://www.cs.ubc.ca/research/flann/uploads/FLANN/flann_visapp09.pdf}

\bibitem{PCL2011}
R.~B. Rusu, S.~Cousins, {3D} is here: Point cloud library ({PCL}), in: 2011
  IEEE International Conference on Robotics and Automation, 2011, pp. 1--4.
\newblock \href {https://doi.org/10.1109/ICRA.2011.5980567}
  {\path{doi:10.1109/ICRA.2011.5980567}}.

\bibitem{Malt2017}
S.~Valat, A.~S. Charif-Rubial, W.~Jalby,
  \href{https://doi.org/10.1145/3141865.3141867}{Malt: a malloc tracker}, in:
  Proceedings of the 4th ACM SIGPLAN International Workshop on Software
  Engineering for Parallel Systems, SEPS 2017, Association for Computing
  Machinery, New York, NY, USA, 2017, p. 1–10.
\newblock \href {https://doi.org/10.1145/3141865.3141867}
  {\path{doi:10.1145/3141865.3141867}}.
\newline\urlprefix\url{https://doi.org/10.1145/3141865.3141867}

\bibitem{geometric-mean}
P.~J. Fleming, J.~J. Wallace, \href{https://doi.org/10.1145/5666.5673}{How not
  to lie with statistics: the correct way to summarize benchmark results},
  Commun. ACM 29~(3) (1986) 218–221.
\newblock \href {https://doi.org/10.1145/5666.5673}
  {\path{doi:10.1145/5666.5673}}.
\newline\urlprefix\url{https://doi.org/10.1145/5666.5673}

\end{thebibliography}
\balance







\end{document}